%% file: Paper.tex
\documentclass[journal]{IEEEtran}
\usepackage[pdftex]{graphicx}
\usepackage[cmex10]{amsmath}
\usepackage{amssymb}
\usepackage{multicol}
\usepackage{colortbl}%
  \newcommand{\myrowcolour}{\rowcolor[gray]{0.925}}
\usepackage{longtable}
\usepackage{caption}
\usepackage{subcaption}
\usepackage{cite}
\usepackage{psfrag}
\usepackage{epstopdf}
\usepackage{algorithm}
\usepackage{algorithmic}
\usepackage{widetext}
\usepackage{float}
\usepackage{color}
\usepackage{soul}
\usepackage{upgreek}
\usepackage{multirow}
\usepackage[justification=centerlast]{caption}
\usepackage{tikz,pgfplots}
\usepackage{dblfloatfix}
\usepackage{textcomp}
\usepackage{adjustbox}
\DeclareUnicodeCharacter{2005}{\hspace{0.01em}}
\usepackage{tikz} 
\usetikzlibrary{calc}
\usepackage{mathtools}
\usetikzlibrary{arrows}
\pgfplotsset{compat=newest}
\usepgfplotslibrary{fillbetween}
\usetikzlibrary{patterns}

\usepackage[hyphens]{url}
\usepackage[hidelinks]{hyperref}
\hypersetup{breaklinks=true}
\usepackage{cite}

\definecolor{myBlue}{RGB}{72,125,215}
\definecolor{myOrange}{RGB}{118,54,45}
\definecolor{InfinBlue}{RGB}{72,72,51}

\ifCLASSINFOpdf
\else
\fi
\begin{document}
%
\title{Implementing Neural Network-Based Equalizers in a Coherent Optical Transmission System Using Field-Programmable Gate Arrays}
     \pgfplotsset{
        compat=1.3, 
        my axis style/.style={
            every axis plot post/.style={/pgf/number format/fixed},
            ybar=5pt,
            bar width=8pt,
            x=1.7cm,
            axis on top,
            enlarge x limits=0.1,
            symbolic x coords={MLP, biLSTM, ESN, CNN+MLP, CNN+biLSTM, DBP},
            visualization depends on=rawy\as\rawy, 
            nodes near coords={%
                \pgfmathprintnumber[precision=2]{\rawy}
            },
            every node near coord/.append style={rotate=90, anchor=west},
            tick label style={font=\footnotesize},
            xtick distance=1,
        },
    }
%

\author{Pedro J. Freire, Sasipim Srivallapanondh, Michael Anderson,  Bernhard Spinnler, Thomas Bex, Tobias A. Eriksson, Antonio Napoli, Wolfgang Schairer, Nelson Costa, Michaela Blott, Sergei K. Turitsyn, Jaroslaw E. Prilepsky
\thanks{This paper was supported by the EU  Horizon 2020 program under the Marie Sklodowska-Curie grant agreement 813144 (REAL-NET) and 956713 (MENTOR). JEP is supported by Leverhulme Trust, Grant No. RP-2018-063. SKT acknowledges support of the EPSRC project TRANSNET}
\thanks{Pedro J. Freire, Sasipim Srivallapanondh, Michael Anderson,  Jaroslaw E. Prilepsky  and Sergei K. Turitsyn are with Aston Institute of Photonic Technologies, Aston University, United Kingdom, p.freiredecarvalhosouza@aston.ac.uk.}
\thanks{Antonio Napoli, Wolfgang Schairer and  Bernhard Spinnler are with Infinera R\&D, Sankt-Martin-Str. 76, 81541, Munich, Germany. anapoli@infinera.com.}
\thanks{Nelson Costa is with Infinera Unipessoal, Lda, Rua da Garagem nº1, 2790-078 Carnaxide, Portugal, ncosta@infinera.com.}
\thanks{Michaela Blott is with AMD Research, Logic Drive, Citywest Business Campus, Saggart, Dublin, Ireland
, michaela.blott@amd.com.}

\thanks{Manuscript received xxx 19, zzz; revised January 11, yyy.}}

%
%

\markboth{Journal of Lightwave technology , ~Vol.~y, No.~x, November~2022}%
{Shell \MakeLowercase{\textit{et al.}}:  Implementing Neural Network-Based Equalizers in a Coherent Optical Transmission System Using Field-Programmable Gate Arrays}
%



\maketitle
\begin{abstract}
In this work, we demonstrate the offline FPGA realization of both recurrent and feedforward neural network (NN)-based equalizers for nonlinearity compensation in coherent optical transmission systems. First, we present a realization pipeline showing the conversion of the models from Python libraries to the FPGA chip synthesis and implementation. Then, we review the main alternatives for the hardware implementation of nonlinear activation functions. The main results are divided into three parts: a performance comparison, an analysis of how activation functions are implemented, and a report on the complexity of the hardware. The performance in Q-factor is presented for the cases of bidirectional long-short-term memory coupled with convolutional NN (biLSTM + CNN) equalizer, CNN equalizer, and standard 1-StpS digital back-propagation (DBP) for the simulation and experiment propagation of a single channel dual-polarization (SC-DP) 16QAM at 34~GBd along 17$\times$70km of LEAF. The biLSTM+CNN equalizer provides a similar result to DBP and a 1.7~dB Q-factor gain compared with the chromatic dispersion compensation baseline in the experimental dataset. After that, we assess the Q-factor and the impact of hardware utilization when approximating the activation functions of NN using Taylor series, piecewise linear, and look-up table (LUT) approximations. We also show how to mitigate the approximation errors with extra training and provide some insights into possible gradient problems in the LUT approximation. Finally, to evaluate the complexity of hardware implementation to achieve 200G and 400G throughput, fixed-point NN-based equalizers with approximated activation functions are developed and implemented in an FPGA. 
\end{abstract}


%
\IEEEpeerreviewmaketitle

\section{Introduction}

\IEEEPARstart{O}ver the previous couple of decades, the race to find various compensation methods to mitigate the nonlinearities of the fiber and components has produced several noticeable high-performance solutions\cite{guiomar2012mitigation,6717017,4738549, 6118297,sorokina2016sparse,sorokina2017ripple,turitsyn2017nonlinear }. However, due to the high complexity of the proposed solutions, only a few published studies \cite{7369081,vasylchenkova2021fixed,8360165, 8452982} have been conducted to implement these solutions in hardware, e.g., in a field programmable gate array (FPGA) or application-specific integrated circuit (ASIC) 
Recently, machine learning (ML)-based techniques have started to penetrate more and more into different digital signal processing (DSP) applications. 
Therefore, it is natural now to consider how nonlinear equalizers may be designed, addressing NN-based setups while simultaneously taking into account the issues of flexibility and computational complexity.  

A significant number of novel NN-based DSP methods have been developed as a result of research on artificial neural networks (NN) for optical channel equalization: these methods can often provide better performance than that rendered by ``conventional'' DSP approaches while maintaining competitive computational complexity in terms of the real multipliers number~\cite{6975096,WANG20171, hager2018nonlinear, 9083434, Bitachon20, 9184798, melek2020nonlinearity, zhang2019field, freire2020complex, schaedlerrecurrent}. However, such investigations typically deal with the software level. In turn, we stress that a few important extra steps are needed to perform a true evaluation of an NN-DSP device at the hardware level and that this creates some new problems and challenges.

In coherent digital optical transceivers, an FPGA is often used as a prototype to assess the performance of an ASIC~\cite{song2022real}. As a result, it is desirable that the NN-based equalizers are implemented in the FPGA to assess the practicability of algorithms used in real-time systems. At this early stage, the FPGA implementation of the NN-based equalizers can also be based on offline processing~\cite{faruk2017digital}. In this paper, we outline the procedures required to move both recurrent and feedforward NN-based equalizers (to be deployed in coherent long-haul optical systems) from the software level (Python) to the FPGA realization. Note that our approach can be applied to all NN architectures for channel equalization, so the aforementioned NN architectures are taken just to exemplify the case. However, it should be noted that our research is also important for the other fields in ML applications, as FPGA-based accelerators have been increasingly attracting interest due to their high performance, energy efficiency, fast development cycle, and reconfiguration capability \cite{zhang2015optimizing}. Furthermore, the driving force behind the deployment of FPGAs is their cloud services applications\cite{nakahara2020high,yen2004fpga}. 

In this paper, we make a step forward in assessing the viability of NN-based equalizers for industrial applications by benchmarking: i) their performance versus the 1-step-per-span (StpS) digital back-propagation (DBP) using 2.3~samples/symbol (Sa/symbol) in experiments, and ii) their computational complexity by comparing an FPGA implementation against the full electronic chromatic dispersion compensation (CDC) block in the time domain implementation (used, e.g., in standard DSP chain~\cite{xu2010chromatic}) that needs much fewer resources than the 1-StpS DBP. In addition, for the first time, to the best of our knowledge, we present the FPGA implementation of an NN-based equalizer that employs the bidirectional recurrent layer with long-short-term memory(LSTM) cells (biLSTM). By transmitting a 34~GBd single-channel, dual-polarization (SC-DP) 16QAM signal over 17$\times$70 km of large-effective area fiber (LEAF) (both simulated and experimental cases), we report $\approx\!1.7$ dB Q-factor improvement over a standard DSP chain while requiring only $\approx\!2.5$ times more FPGA resources than the implementation of the CDC block to achieve a 400G transmission.

This paper is organized as follows. Sec.~\ref{sec:review} reviews the previous implementations of NN structures in FPGA, with a special focus on their application for the optical channel equalization task. Sec.~\ref{sec:pipeline} describes the steps taken to create the NN-based equalizer, from software to hardware. In this section, we introduce the 4 steps in our realization pipeline using the Xilinx tools for high-level synthesis (Vitis HLS) and hardware synthesis (Vivado\textregistered). Sec.~\ref{sec:activation_funtion} presents a complete study on the realization of nonlinear activation functions in hardware using the three most common approximators: the Taylor approximator, the piecewise linear (PWL) approximator, and the look-up table (LUT) approximator. In this section, the drawbacks of using each of these approximation techniques for performance and complexity are shown. Sec.~\ref{sec:results} describes the experimental and simulated setups used and the performance in terms of Q-factor for both simulation and experimental datasets. This section also talks about how well different approaches to approximating the activation functions work and how much hardware they use. Finally, we report the computational complexity (utilization), latency, and throughput for all NN strategies studied in our manuscript versus respective quantities of the CDC block when using all available resources on the FPGA under investigation (VCK190~\cite{FPGA}) and when using only LUT and flip-flops (FF) to simulate a realization closer to the ASIC. The last section concludes our paper with a summary of our approach, the results achieved, and some open questions in this field.

\section{FPGA Designs for ML-based Equalization in Optical Transmission} \label{sec:review}
The FPGA is a programmable and reprogrammable integrated circuit that is suitable for resource-constrained embedded applications, as it provides more energy-efficient computation when performing NN on the edge compared to the GPU \cite{pettersson2020convolutional, nurvitadhi2016accelerating}. FPGA implementations have been investigated for different types of NN, for both feedforward \cite{pettersson2020convolutional,qiu2016going, hajduk2018reconfigurable} and recurrent NNs (RNN), including LSTM \cite{nurvitadhi2016accelerating, chang2015recurrent,huang2017implementation,chang2017hardware}. As NNs can be used in numerous areas, the FPGA design for NN has been intensively researched in different applications: signal processing \cite{dhavlle2020comprehensive}, industrial control applications \cite{monmasson2011fpgas}, drive systems \cite{orlowska2011fpga}, and telecommunication equalization \cite{kaneda2020fpga,kaneda2021fixed,huang2021recurrent,li2021fpga,giacoumidis2020real,yen2004fpga,chen2005fpga,lee2020neural}.

Among the very first works discussing the NN-based equalizers in FPGA, Yen et al. \cite{yen2004fpga} proposed the functional link artificial NN (FLANN) for nonlinear channel equalizers in both software simulation and hardware implementation in FPGA, considering the quadrature phase shift keying (QPSK) modulation. The FLANN was claimed to have a simpler architecture and higher computational speed in hardware compared to the multi-layer perceptron (MLP). Performance comparison was carried out to compare FLANN with the linear least-mean squares equalizer (LMSE). FLANN provided a better symbol error rate than LMSE. However, with parallel processing, FLANN required more logic cells and more area of the chip. To reduce this, the number of data bits in the decimal fraction should be reduced with a trade-off in system performance.
Subsequently, the nonlinear channel equalizer based on the NN radial basis function (RBF) with a three-layer structure implemented on FPGA, was investigated in~\cite{chen2005fpga}. The results indicated that the bit error rate (BER) performance in the software simulation and that of the Bayesian equalizer, which is a near-optimal method of a channel equalizer, were similar. However, the hardware implementation showed worse results due to the binary value representation on the hardware. To approach the BER of the original RBF NN structure, an increase in the number of bits was recommended. 

In more recent works, the convolutional NN (CNN) and binary CNN-based decision schemes for millimeter-wave (mm-wave) radio-over-fiber (RoF) optical communications were presented \cite{lee2020neural}. Such NN structures outperformed the MLP\footnote{Note that in~\cite{lee2020neural}, the MLP is referred to as the fully-connected NN (FCNN).}-based equalizer in terms of complexity, but still achieved the BER performance within the forward error correction (FEC) limit when tested with the 60~GHz mm-wave RoF system. The FPGA-based CNN and binary CNN hardware accelerators with inner parallel optimization were implemented to verify the capability of the FPGA compared to the GPU. Their results showed that the FPGA-based hardware can significantly reduce latency, cost, and power consumption while demonstrating comparable performance. 
\begin{figure*}[th!]
\centering
\begin{subfigure}{.5\textwidth}
  \centering
  \includegraphics[width=\linewidth]{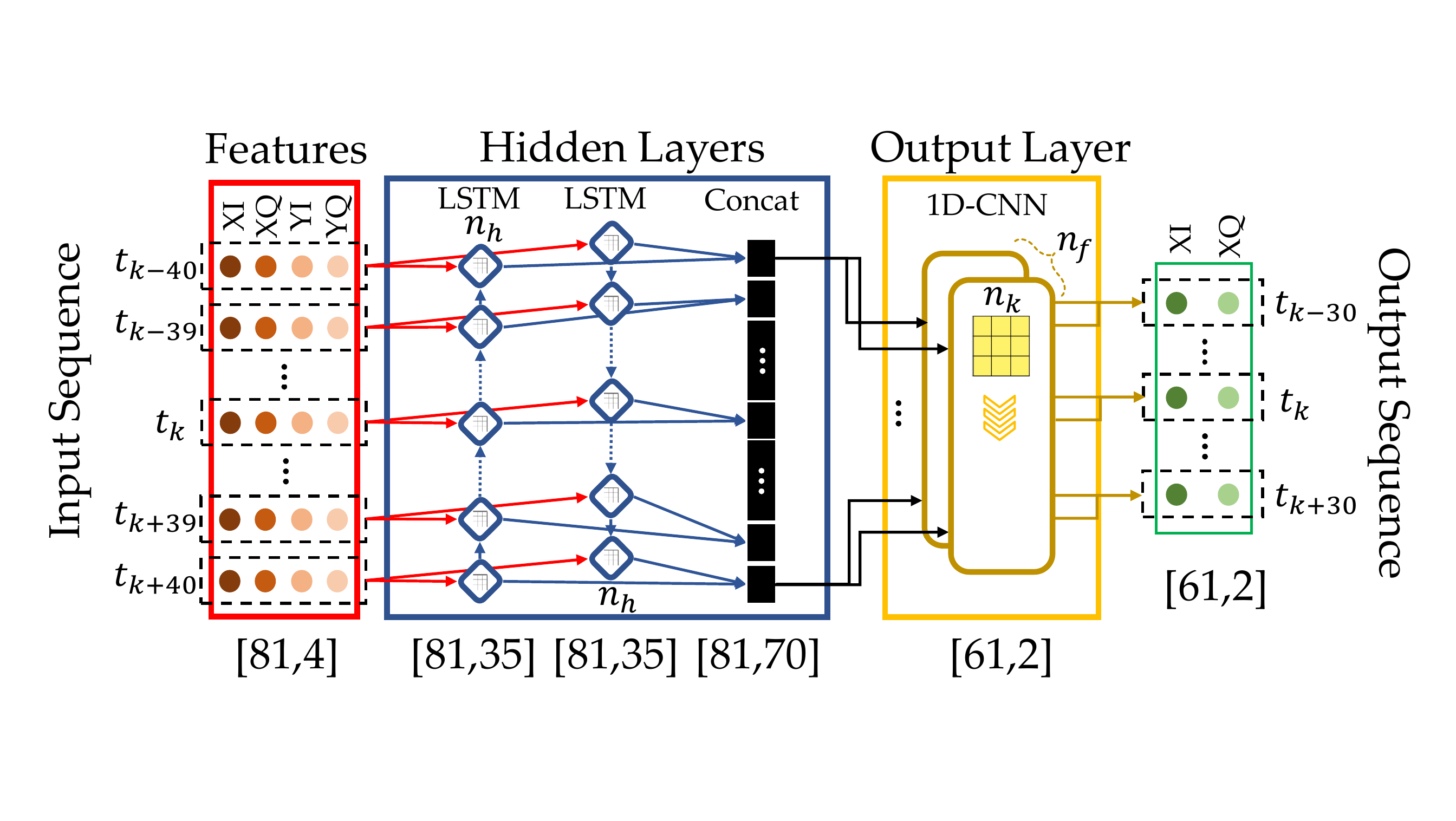}
    \vspace{-10mm}
  \caption{biLSTM+CNN.}
  \label{fig:nn_structure_lstm}
\end{subfigure}%
\begin{subfigure}{.5\textwidth}
  \centering
  \includegraphics[width=\linewidth]{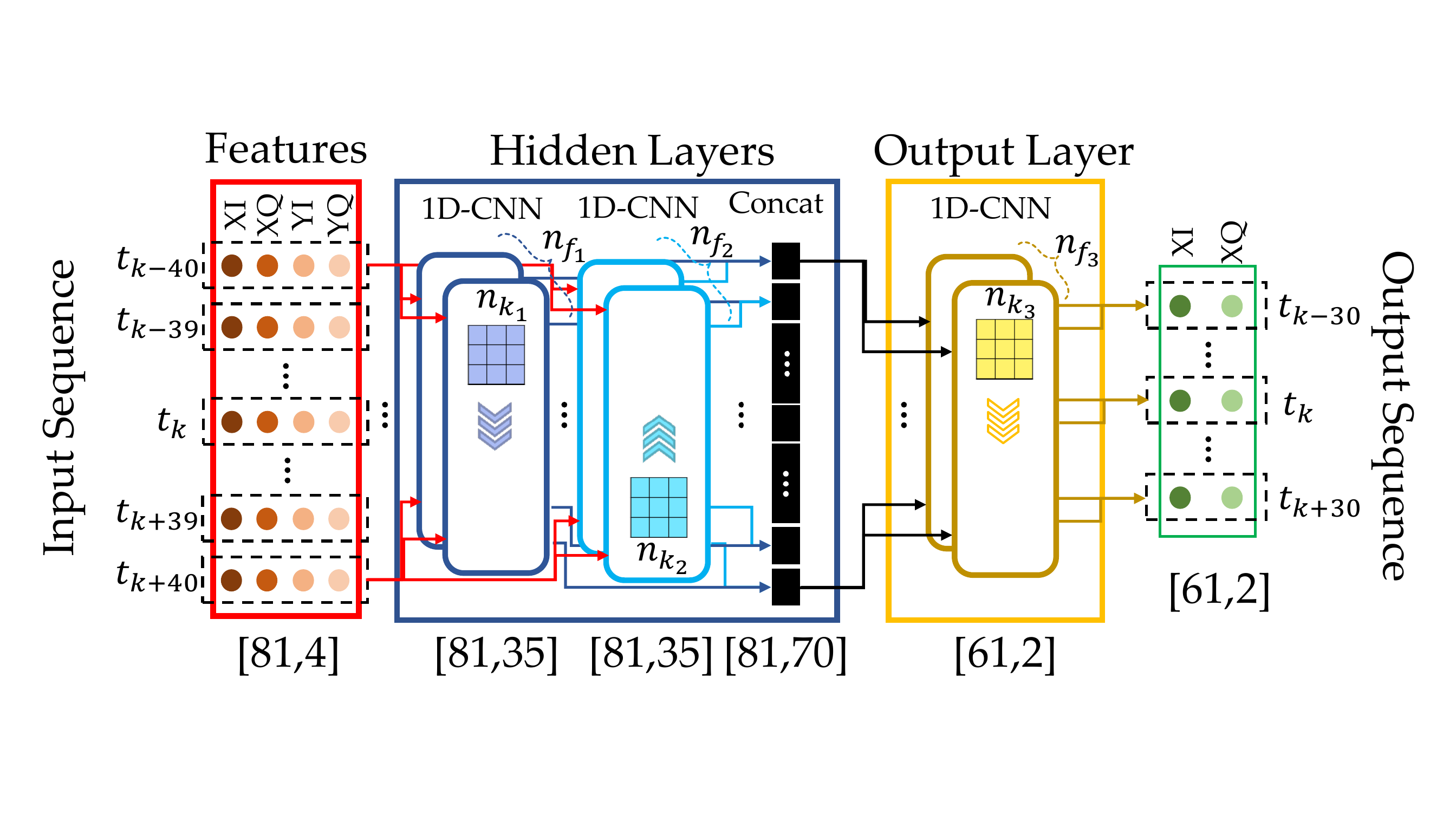}
    \vspace{-10mm}
  \caption{Deep CNN.}
  \label{fig:nn_structure_cnn}
\end{subfigure}
\vspace{-5mm}
\caption{Structures of NN-based equalizers taking 81 symbols as input to recover 61 symbols in parallel at the output: (a) the recurrent equalizer using a bidirectional LSTM layer containing 35 hidden units, and (b) the feedforward equalizer using a 1D-convolutional layer consisting of 70 filters ($n_{f_1}=n_{f_2}=35$).}
\label{fig:nn_structure}
\end{figure*}

The studies of the implementation of FPGA-based optical equalizers based on NN have recently gained additional attention, but mainly for direct detection systems \cite{kaneda2020fpga, huang2021recurrent, li2021fpga}. In~\cite{kaneda2020fpga}, the MLP-based equalizer contained two hidden layers with 33 and 14 neurons, respectively, to equalize the 50 Gb/s passive optical networks (PONs) link. The authors implemented an 8-bit fixed point deep NNs in an FPGA and showed that deep NNs with embedded parallelization successfully reduced the required hardware resources. The authors have extended their work and reported, in detail, in \cite{kaneda2021fixed}, the impact of fixed-point resolution on receiver sensitivity and the utilization of hardware resources in the FPGA implementation of DNN equalizers for PON systems.

The parallel output RNN-based equalizer was proposed in \cite{huang2021recurrent}. This parallel RNN is superior to the parallel MLP demonstrated by \cite{kaneda2020fpga} in terms of BER, as was shown for the 100~Gbps/$\lambda$ PON. However, the feedback loop structure in the RNN caused hardware implementation challenges, as the output of the n$^{th}$ time step has to propagate back in the loop and appear at the input before the neuron begins to process the n+1$^{th}$ input. As a result, the authors of that reference only evaluated the parallel MLP. 
In \cite{li2021fpga}, the time-interleaved parallel pruned MLP-based equalizer for 100~Gbps PAM-4 links was implemented on an FPGA. The NN structure has three layers, and the hidden layer contains 51 neurons. The reported weight pruning algorithm, in this work, is a novel pruning algorithm based on weighting probability to reduce computational complexity while maintaining performance. They reported that a single pruned NN-based equalizer achieved over 55\% resource reduction compared to the NN before pruning. Furthermore, up to 40\% of resource utilization was reduced for 4- and 8-channel equalization.

\begin{figure*}[th!]
      \centering
  \includegraphics[scale=0.4]{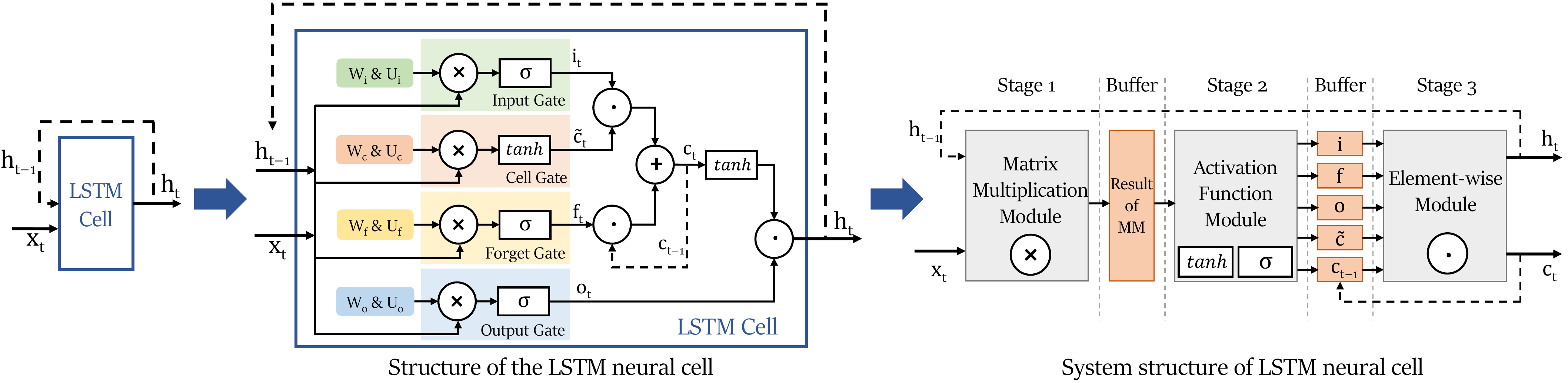}
  \caption{LSTM cell structure showing the recursive connections (detailed in Section~\ref{subsec:NN-pyton}) and the system structure of LSTM implementation in a modular way showing the buffers needed to store outputs (detailed in Section~\ref{sec:CCvsCDC}). }
  \label{fig:lstmcell}
\end{figure*}

However, work on the FPGA implementation of NN-based equalizers in the case of coherent detection optical systems is still mostly missing. We mention only~\cite{giacoumidis2020real}, where the authors demonstrated the mitigation of optical fiber non-linearity in a 16-QAM self-coherent real-time system at 40 Gb/s using a FPGA implementing the sparse \texttt{K-means++} algorithm instead of an NN. In this case, the authors reported a 3 dB Q-factor improvement with respect to linear equalization only after transmission along 50~km of optical fiber using a launch power close to the optimal value of 14 dBm. In contrast to the case considered in this work, the tested scenario was a single-span short-reach system.

In our work, we describe and detail the next step in the implementation of NN equalizers: For the first time, the offline FPGA implementation\footnote{Offline FPGA implementation refers to the process of designing and configuring an FPGA before it is deployed in a target system. In this work, we simulate “offline” the constraints affecting an FPGA when taking into account the NN simulation and tested on the VCK190 board with some sample data that were saved in the FPGA's memory just to verify that the model (bitstream file) was working according to its design.} of an NN equalizer employing the recurrent layer (biLSTM), as well as a deep CNN structure, is presented and evaluated in the experimental data for a high-speed coherent optical transmission system.

\section{Neural network equalizers designing pipeline: From Python to FPGA} \label{sec:pipeline}

\begin{figure*}
    \centering
    \includegraphics[width=\linewidth]{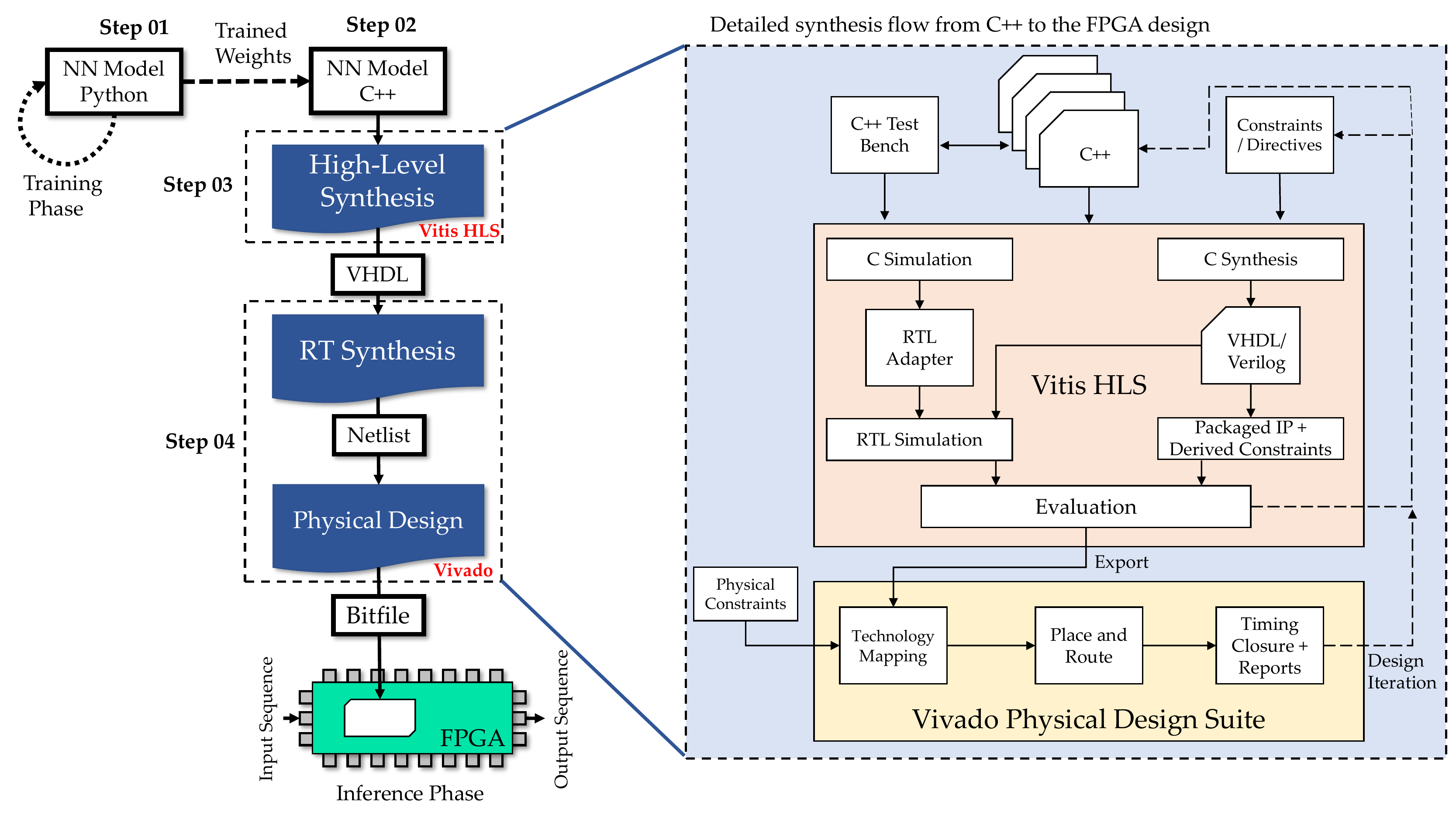}
    \caption{High-level synthesis flow -- from C++ to FPGA realization.}
    \label{fig:FPGA_pipeline}
\end{figure*}

In this section, we look at the design tools and process steps that were used to implement the NN in an FPGA.
In subsection \ref{subsec:NN-pyton}, corresponding to Step 01 in Fig.~\ref{fig:FPGA_pipeline}, the NN architectures studied in this work are presented and the details of the training phase are depicted. In subsection \ref{subsec:C++}, specific attention is devoted to the C++ model (Step 02 in Fig.~\ref{fig:FPGA_pipeline}) and the high-level synthesis (HLS) process (Step 03 in Fig.~\ref{fig:FPGA_pipeline}) used to generate a description of the NN in VHDL (Very High-Speed Integrated Circuit Hardware Description Language). We also explain the motivation behind the use of the HLS method. In addition, we discuss some important considerations for using HLS, intending to support future research activities that involve this method. Then, in subsection \ref{subsec:vivaldo}, we look at the physical implementation aspects of the design flow, performed there by using the Vivado\textregistered\! design suite as shown in Step 04 of Fig.~\ref{fig:FPGA_pipeline}, which produces the final results related to the FPGA hardware.

\subsection{Neural Network Architectures and Python Training Process}\label{subsec:NN-pyton}

The two NN architectures for the equalizers investigated in our work are depicted in Figs.~\ref{fig:nn_structure_lstm} and \ref{fig:nn_structure_cnn} for the biLSTM-based equalizer and the deep CNN-based equalizer, respectively. The shape of both architectures is similar, but the nature of the mathematical operations in each case is different: biLSTM contains recursive connections, as can be seen in Fig. \ref{fig:lstmcell} for the LSTM cell structure, whereas deep CNN is a feedforward network. In Fig. \ref{fig:lstmcell}, the dashed line indicates that the recursive connections due to the equations of a forward pass of an LSTM cell with a time step $t$ are given as:
\begin{equation}\label{eq.lstm}
    \begin{gathered}
i_{t} = \sigma(W_{i}{x}_{t} + U_{i}{h}_{t-1} + b_{i} ),  \\
    f_{t} = \sigma(W_{f}{x}_{t} + U_{f}{h}_{t-1} + b_{f}), \\
o_{t} = \sigma(W_{o}{x}_{t} + U_{o}{h}_{t-1} + b_{o}),\\
\Tilde{c_t} = \phi(W_{c}{x}_{t} + U_{c}{h}_{t-1}+ b_{c})\\
    c_{t} = f_{t}\odot c_{t-1} + i_{t}\odot \Tilde{c_t}, \\
    h_{t} = o_{t} \odot \phi(c_{t}),
    \end{gathered}
\end{equation}
where $\phi$ is usually the ``tanh'' activation function, $\sigma$ is usually the sigmoid activation function, $x_{t}$ is the input vector at time $t$, $W$ and $U$ representing the trainable weight matrices, and $b$ is the bias vector.  $i_t, f_t, o_t, \Tilde{c_t}$, $c_t$, and $h_t$ denote input gate, forget gate, output gate, cell input, cell state, and hidden state vectors, respectively. The $\odot$ symbol represents the element-wise (Hadamard) multiplication. 

The NN-based equalizer is applied after the standard DSP chain\footnote{Hence, the time recovery and other typical DSP blocks for coherent transmission are already handled, and the sampling rate of 1 sample per symbol is already in place.}. In both the biLSTM and deep CNN equalizers, a total of 81 symbols are used as input, allowing us to simultaneously retrieve 61 symbols at the equalizer's output. By recovering 61 symbols in parallel at the equalizer output, we allow the FPGA realization to have a higher throughput, which is one of the main desirable design features we are looking for when building a DSP block for coherent transmission systems. It is worth noting that the NN output layer in this scenario recovers the X polarization, as the CDC block is applied independently to each polarization. However, in future research and implementation, the NN's output layer can also be modified to recover both polarizations simultaneously. In Fig.~\ref{fig:nn_structure_lstm}, the hidden layer consists of a biLSTM layer with $n_h=35$ hidden units. In Fig.~\ref{fig:nn_structure_cnn}, the hidden layer is made up of a CNN layer with 70 filters ($n_{f_1}=n_{f_2}=35$), with zero padding applied to retain the shape and the kernel size $n_{k_1}= n_{k_2}=11$. The output layer in both designs is a convolutional layer with $n_f=2$ filters, a kernel size $n_k =21$, and no padding. Based on a grid search analysis, these parameters were chosen to meet the hardware limitations, throughput requirements, and optical performance required for this FPGA realization. The activation functions for both hidden layers were hyperbolic tangent ($tanh$), and the output layer is linear.

Focusing now on the biLSTM+CNN architecture implemented in this work, the mean square error (MSE) loss estimator and the classical Adam algorithm for the stochastic optimization step~\cite{gulli2017deep} were used when training the weights and bias of the NN. The training hyperparameters (mini-batch size equal to $2001$ and learning rate equal to $0.0005$) were found using the Bayesian optimization procedure described in~\cite{freire2020complex}. The NN training was carried out by backpropagation for 30000 epochs with a fixed set of hyperparameters\footnote{For applying the model to other launch power values, we used transfer learning~\cite{freire2021transfer}, which allowed us to utilize less than 5 epochs to adjust the NN weights to the other launch powers.}. The BER is evaluated after each training epoch. For training, we used a fixed dataset with $2^{20}$ symbols, and, at every epoch, we picked $2^{18}$ random input symbols from this dataset. For the testing and validation, we used a never-before-seen dataset with $2^{18}$ symbols. Both NN models were trained, validated, and tested using the same datasets. The weights were saved at the epoch at which the BER measured using the validation dataset was the lowest (early stopping).

\subsection{C++ Implementation of Neural Networks and the High-Level Synthesis Method to Generate VHDL}\label{subsec:C++}

The HLS method provides a design flow in which the desired function, i.e., an NN in the case considered, can be specified in C++ and then automatically converted to VHDL. Note that the NN implementation in C++ is coded from scratch.
This strategy is preferred because it separates the FPGA technology-specific implementation features, such as clocks and logic cell topology, from the NN's functionality. Working at a higher level of abstraction, the intended functionality can be the focus of attention and be described more easily in fewer lines of code\cite{caba2021towards}. 

In addition, functional verification in C++ is much faster than functional simulation in VHDL. This makes it easy to test and debug the design. At this point, it is important to remember that the functions described will run on hardware. HLS supports a substantial range of C++ syntax, but not all, because some cannot be implemented in an FPGA or ASIC. In this case, memory allocation is a key part of writing the C++ code that needs to be properly assessed. In the FPGA, the memory allocations are static and are assigned during the mapping phase of the physical design flow. Dynamic memory allocations found in many C++ standard library functions cannot be supported and must be avoided or replaced with structures optimized for the implementation in an FPGA by using the libraries provided by the HLS tool supplier, in our case, Xilinx. Also, Operating System (OS) functions such as file read/write and date/time cannot be implemented in the FPGA; all data transmitted into and out of the FPGA must use input/output ports.

Consequently, two versions of the C++ codes were generated. The first is called here the test bench, while the second is the function to be implemented in FPGA. The test bench function reads the previously saved signal inputs and weights learned in Python and converts them to a fixed-point format (int32). These values are then incorporated into the function that describes the equalizers investigated in this study. The function is the C++ translation of the Python NN equalization architecture, using fixed-point arithmetic operations. After the equalization, the outputs of the function (the signal that has been equalized) are delivered to the test bench code that checks the MSE. We did not study the impact of further reducing the quantization accuracy, since this would require some further work to overcome the quantization error caused by both the input signal and the weights. We chose the int32 format because, by using it, we can take advantage of simplified integer arithmetic while observing no significant performance reduction compared to the floating-point BER evaluated in Python. Note that INT32 is the quantization format for input and weights in this paper, and different types of quantization are studied in \cite{freire2023reducing}.

Here, it is important to highlight details on the implementation of the convolutions and the LSTM cells in our C++ code. For the convolution implementation, we used the conventional method for convolution, which consists of a series of for loops that can be partialized, as detailed in Ref.~\cite{baptista2019platform}. In the case of the LSTM cell, l, whose implementation is depicted in Fig. \ref{fig:lstmcell}, our approach adhered to the methods outlined in Ref.~\cite{he2021fpga}. 
In summary, the input data  $h_{t-1}$, $x_t$, and weight matrix $W$ are read, and the systolic array technique is used to do the matrix multiplication. Its output is temporarily stored in global memory on the chip. Then, the activation function module (e.g., Taylor approximation, PWL approximation, or LUT approximation) receives the input data from the temporary result buffer and obtains the output vectors $i$, $f$, $o$, and $g$, as shown in equation 1. Each gate's output is also buffered and saved in the chip's global memory. Next, we implemented the element-wise computation module, which reads the data of $i$, $f$, $o$ and $g$ from the buffer, completes the element-wise computation, as shown in Fig.~\ref{fig:lstmcell} and then obtains the output $h_t$ and cell state $c_t$, which will be used in the next time step. After all required time steps have been completed, the final output is written back to the host memory.

However, when using the HLS, even though the conversion to VHDL is automated, some design intervention is still necessary, and engineering decisions must be taken to achieve the desired performance. HLS supports a set of directives, or pragmas, that can be used to modify the behavior of the HLS C++-synthesis stage to facilitate these interventions \cite{documentation_portal_2022}. By utilizing pragmas, in order to discover the best implementation, it is useful to investigate several design structures without re-coding them. Although there are a variety of different pragmas, we have primarily utilized those pertaining to pipelines, loops, and arrays.

Pipelines allow the parallel execution of operations within a function, lowering the number of clock cycles between commencing loop iterations; these clock cycles are referred to as the Iteration Interval (II). Each loop iteration does not need to end before the next one begins, i.e., the iterations can overlap. The number of pipeline stages can be controlled by setting the value of II using the HLS pipeline pragma. Setting the II to 1, as is done in this project, enables each cycle to begin with a new iteration.

Loops can be unrolled and flattened. By default, loops within a function remain rolled, which means that the loop body is executed sequentially, utilizing a single set of logic resources. The minimum loop delay is then equal to the number of loop iterations. Unrolling generates several copies of the loop body logic, enabling parallel execution and reduced latency at the expense of additional size. Loops can be entirely or partially unrolled, resulting in either one copy of the loop body every iteration for optimum throughput or fewer copies for reduced area cost. Flattening transforms a hierarchy of nested loops into a single loop, which eliminates a clock cycle delay while traveling between higher and lower nested loops and can help with better optimization of the loop logic. Given this, the loops in the feedforward NN layers can be flattened; however, the loops in the recurrent NN layers cannot be flattened due to their memory dependence and imperfect loops. Therefore, for the implementation of the recurrent NN layer, we shall only flatten the loops relating to the matrix multiplication that occurs internally within each LSTM cell.

Arrays can be partitioned and reshaped. HLS will implement arrays in the C++ code as memory blocks in the VHDL description. This can cause a restriction in the design concurrency as FPGA memory blocks only have 2 access ports, which, if the designer has also used the previously mentioned unroll and pipeline pragmas, will need to be shared between all instances of the loop body. By reshaping and partitioning the array, the size, and the number of these memory blocks can be controlled. To enable the successful flattening of the loops in each NN architecture, the non-equalized signal (input signal) and the weights of the NN architecture are partitioned here. The final stage of the HLS step is to export the generated VHDL and derived constraints for use in the physical design step.

Here, we emphasize that NNs are an excellent candidate for exploiting the benefits of HLS, as their nested architecture consisting of multiple layers and several multiply/accumulate functions can make good use of the loop and pipeline directives to investigate the trade-off between area and latency to meet the design requirements. 

\subsection{Vivado and the Synthesis Step to the FPGA Realization}\label{subsec:vivaldo}

The area and timing reports generated by the HLS stage are still simply estimates of the final design performance based on the technology-specific data libraries for each FPGA; the actual performance cannot be determined until the physical implementation is complete. In our work, the Vivado Design suite from Xilinx is utilized. The physical implementation is performed by Vivado in three steps: technology mapping, placement and routing, and timing analysis.

\textbf{Technology Mapping}. Within this step, the VHDL source code is translated into primitive logic gates and boolean equations, followed by mapping these gates onto FPGA customizable logic blocks containing D-type FFs (DFFs) and RAM-based LUTs or more specialized functional cells, such as DSPs. During the technology mapping, the design is optimized and unnecessary logic is eliminated. Note that the detailed explanation of the FPGA components can be found in Appendix \ref{app:fpga}.

The next two steps constitute an iterative process executed automatically by the tool based on design constraints, such as a clock frequency. These limitations can be inherited from the HLS stage or defined in Vivado. The Vivado tool imposes sets of restrictions through established optimization strategies, which are discussed in detail in the vendor user manuals\cite{documentation_portal_2022_2}, and which the user can apply depending on the design goals. The optimal technique is determined by balancing computer runtime and outcomes. In~\cite{miscircuitos_2021} we can find all potential pairings of synthesis and optimization procedures in Vivado using a high-speed pulse width modulation circuit as a target design, as well as a comprehensive evaluation of the runtime versus performance of the different Vivado optimization methodologies. Since the goal of our work was to increase throughput, we did not look at solutions that would reduce chip size, power, or runtime. Therefore, the Vivado configuration called ``Performance ExtraTimingOpt'' is adopted in our work, since it effectively optimizes throughput by reducing timing slack\cite{miscircuitos_2021}.

\textbf{Placement and Routing}. This stage positions the logic blocks developed during the mapping phase, onto the specified elements of the FPGA cell array, and configures the signal routing channels between them. The placement algorithm starts from a random seed position and then moves functions to the cell array based on the degree of congestion for the parts of the die and the fanout of the driving function.

\textbf{Time Analysis}. This stage compares the design with the applied timing constraints to determine whether the overall performance requirement has been met. Timing analysis, in particular, requires a grasp of the FPGA structure and how the design has been mapped onto the array. It may be necessary to return to the HLS phase to apply more directives or adjust the function architecture to achieve timing closure. The time for a data (signal) to travel between two points is determined by a variety of factors, including DFF switching time, setup requirements (the time at which the signal must arrive at the destination before the capturing clock edge), logic and routing path delays, and clock edge uncertainty due to jitter and clock path skew. Here, it is pertinent to define the negative timing slack.

The negative timing slack indicates that the total delay in the data path between two DFFs is greater than the requested clock period. In this negative slack case, the NN has many nested loops, as discussed in the previous section; unrolling these loops would lead to a larger design consuming more logic area, but leaving large loops, i.e., the loops with a high number of iterations, can produce long logic multiplexer paths as the inputs to the loop logic are selected. Using Vivado timing analysis reports and annotated netlist viewer, we can identify which paths can be the cause of the failed paths. In this case, the solution was to return to the C++ source and reorder the nested loops so that the outer loop, which was not unrolled, had fewer iterations; this approach reduced the size of the logic chain in the multiplexer path. In this work, we have also decreased the clock frequency for each of the designed blocks to guarantee that a zero negative timing slack was achieved in all FPGA designs.

\section{The nonlinear activation function implementation: high accuracy and lower complexity} \label{sec:activation_funtion}

\begin{figure}[t]
    \centering
    \includegraphics[scale=0.3]{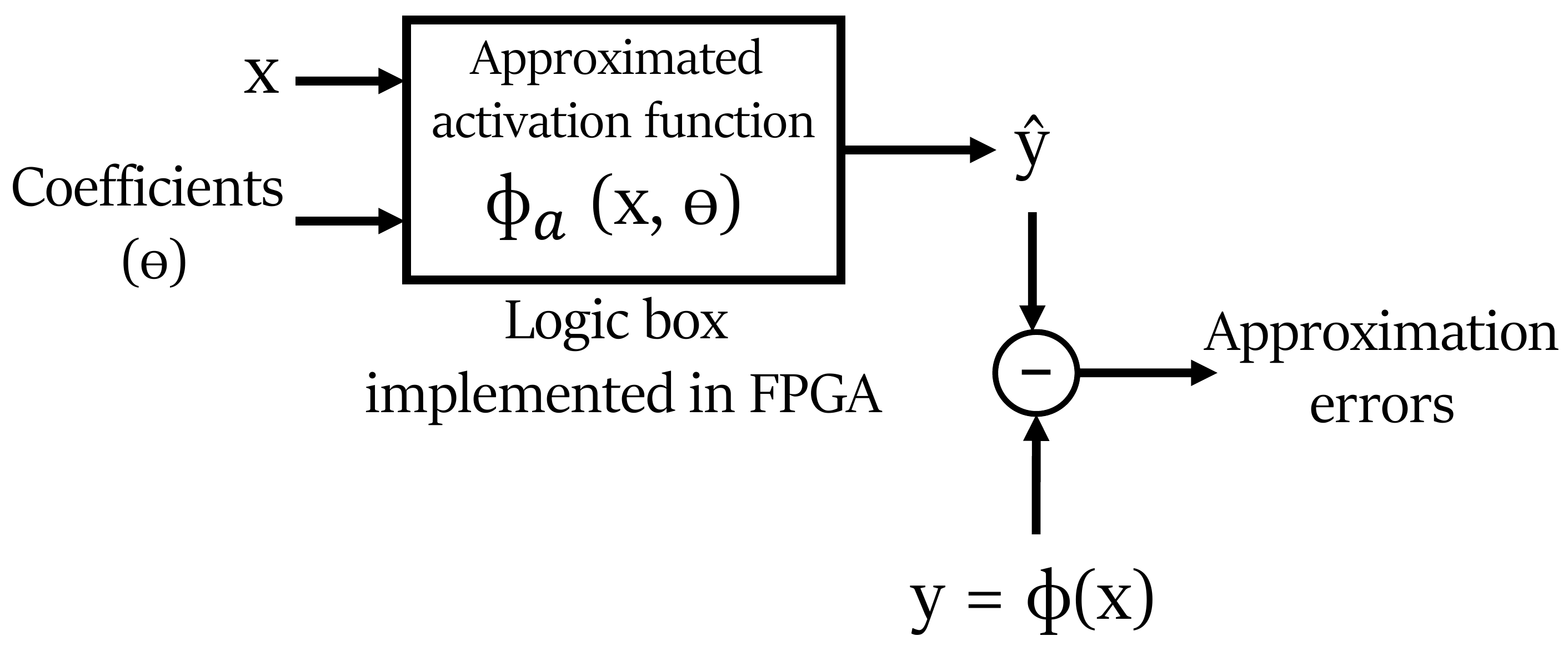}
    \caption{Diagram of the input/output of approximated activation functions based on the logic box implemented in FPGA.}
    \label{fig:acti_func_fpga}
\end{figure}

 \begin{figure*}[t]
\begin{minipage}{.48\textwidth}
\centering
\subcaptionbox{Tanh up to 3\textsuperscript{rd} order.\label{fig:taylor3_tanh}}{\includegraphics[width=.46\linewidth]{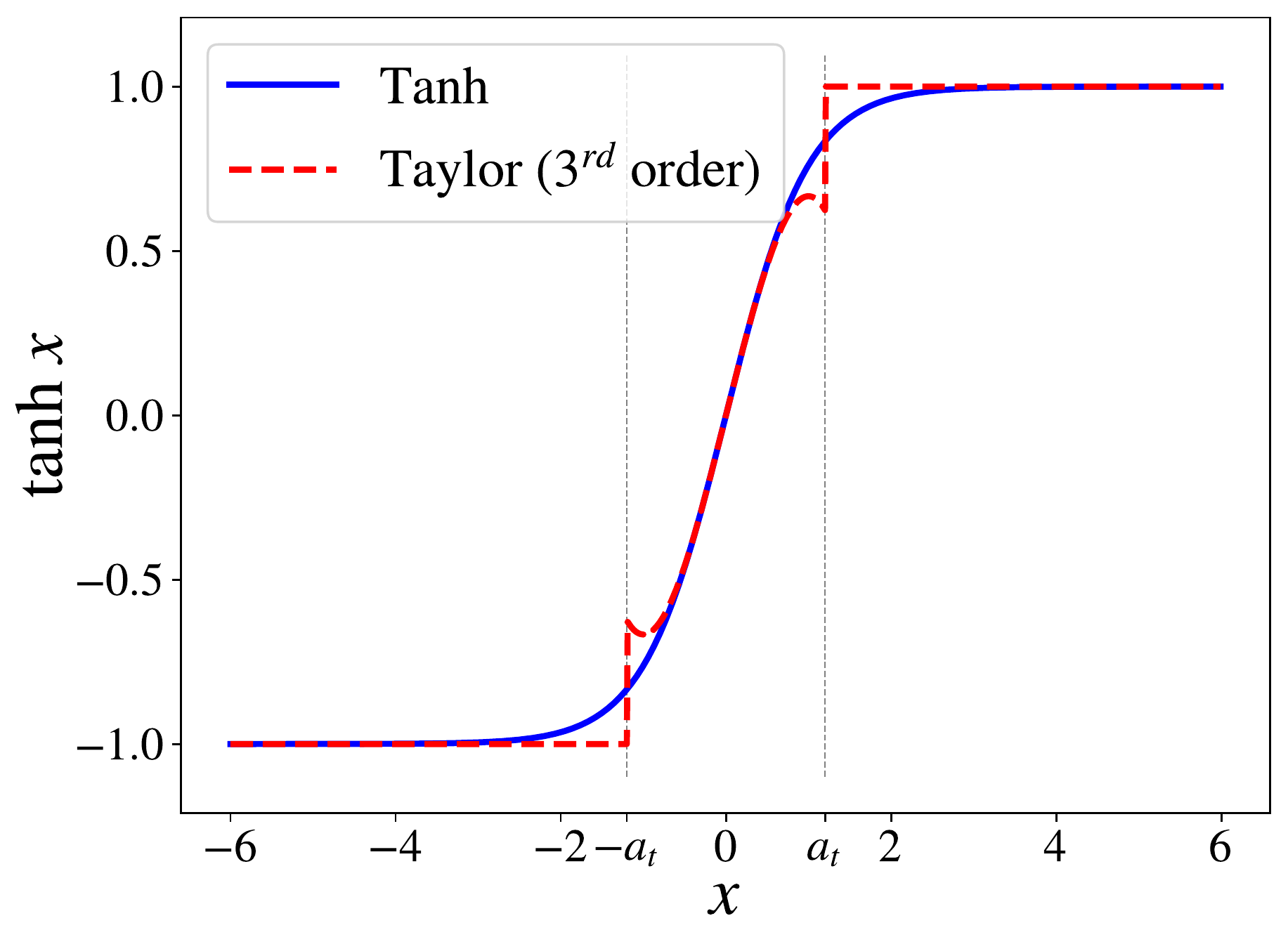}}\quad
\subcaptionbox{Tanh up to 9\textsuperscript{th} order.\label{fig:taylor9_tanh}}{\includegraphics[width=.46\linewidth]{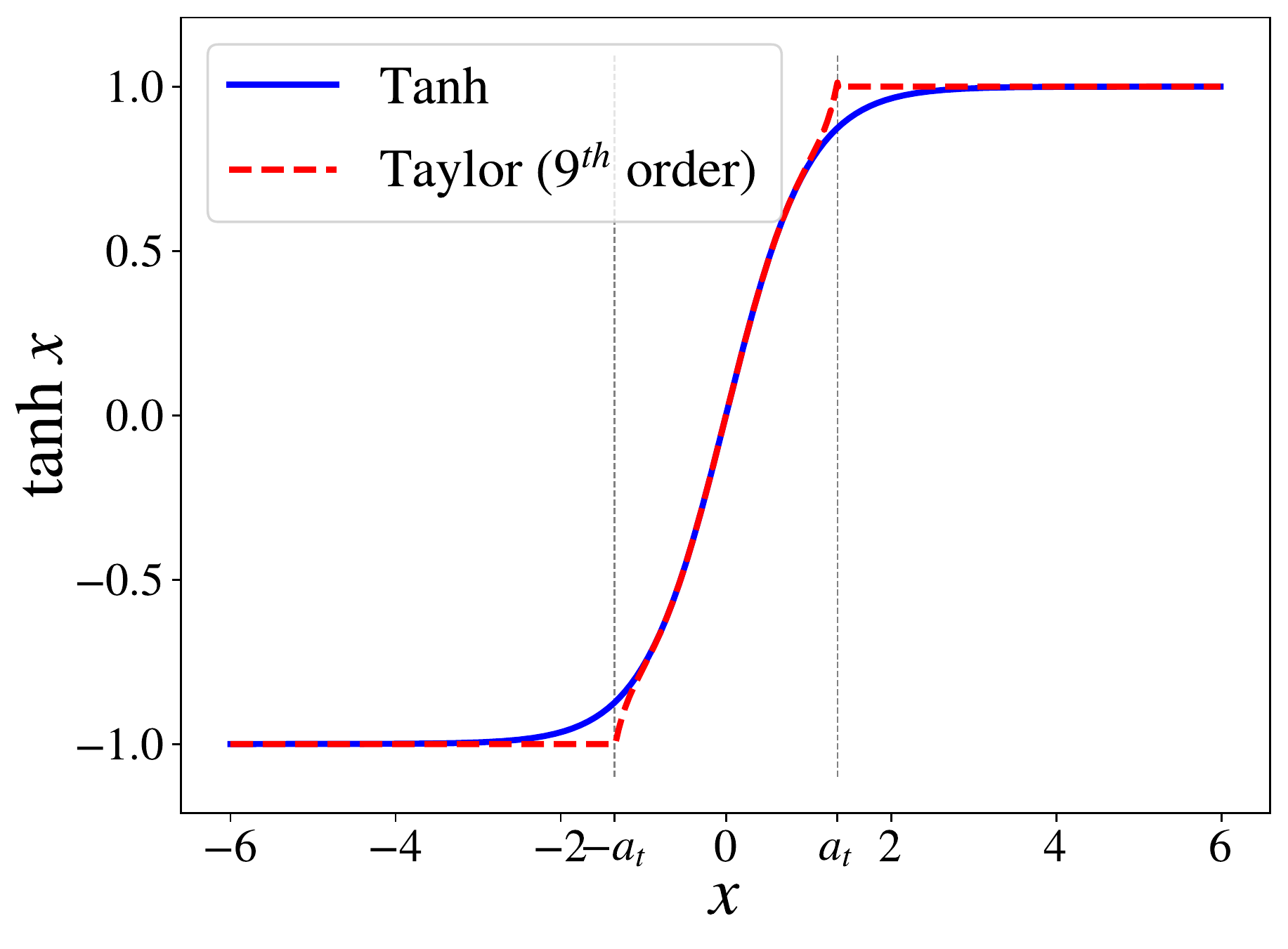}}
\medskip
\vspace{3mm}
\subcaptionbox{Sigmoid up to 3\textsuperscript{rd} order.\label{fig:taylor3_sigmoid}}{\includegraphics[width=.47\linewidth]{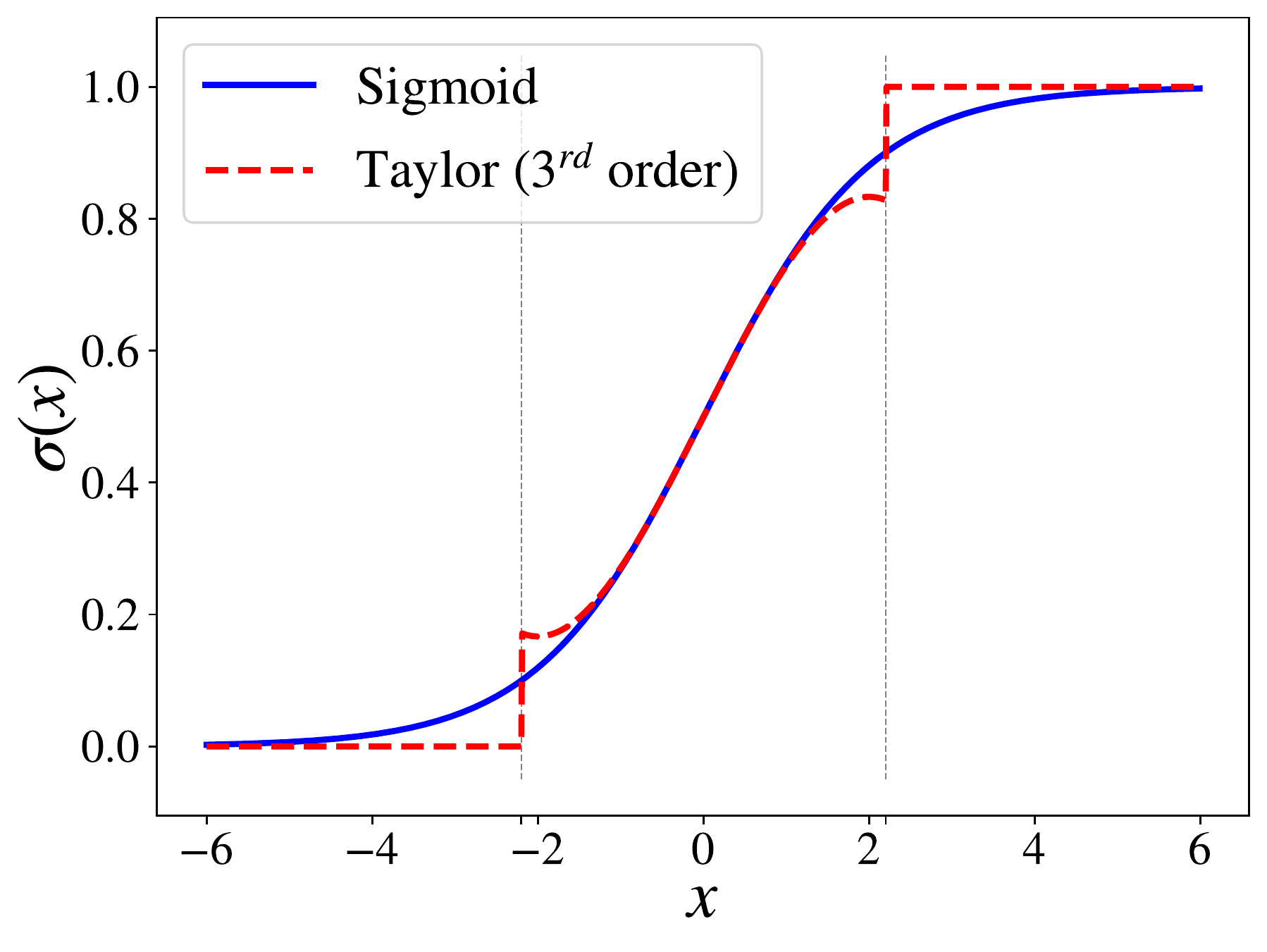}}\quad
\subcaptionbox{Sigmoid up to 9\textsuperscript{th} order.\label{fig:taylor9_sigmoid}}{\includegraphics[width=.47\linewidth]{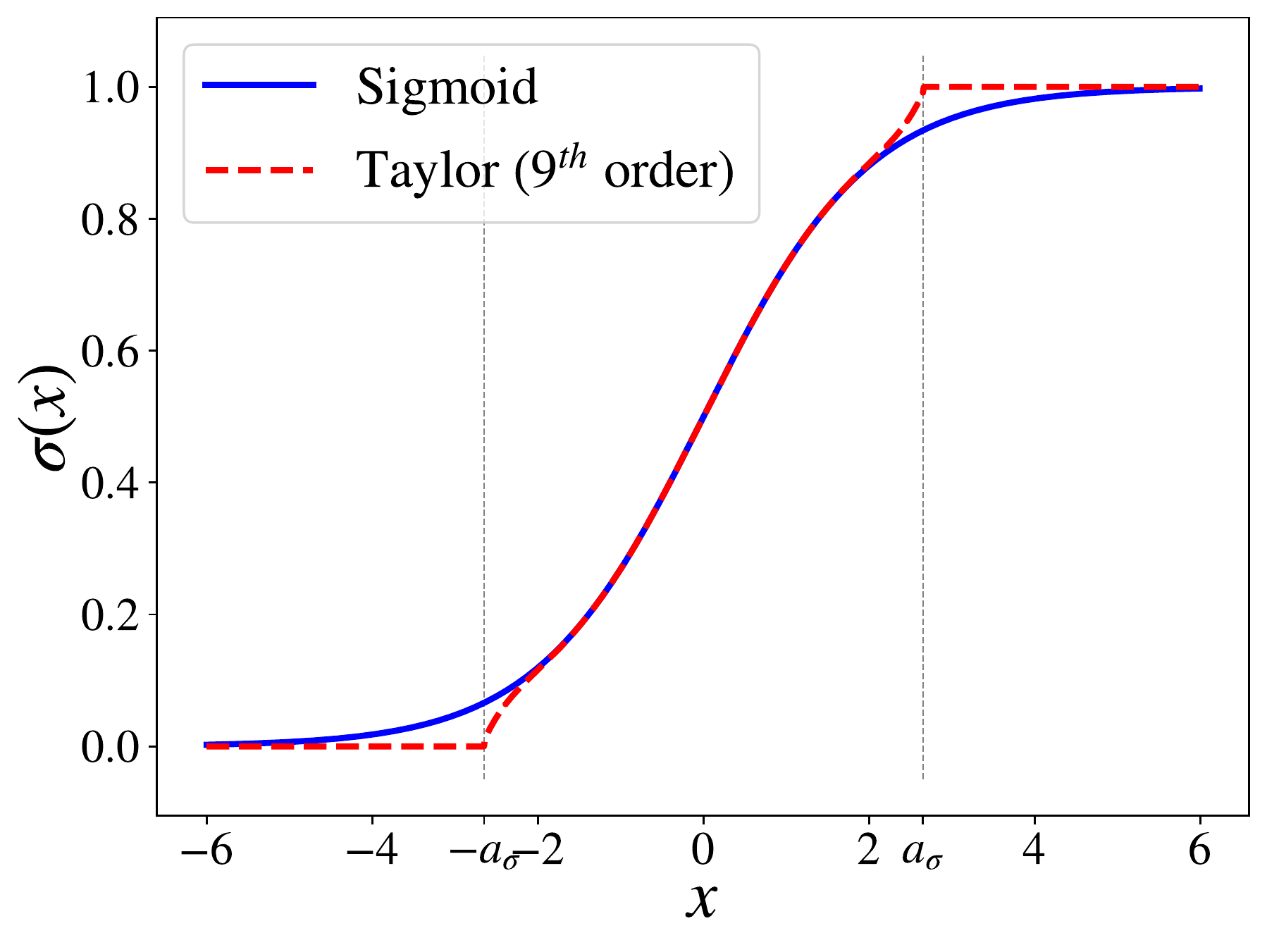}}
\caption{Taylor series approximation of tanh (a) -- (b) and sigmoid functions (c) -- (d).}
\label{fig:taylor}
\end{minipage}\hfill
\begin{minipage}{.48\textwidth}
\centering
\subcaptionbox{Tanh with 3 segments.}{\includegraphics[width=.47\linewidth]{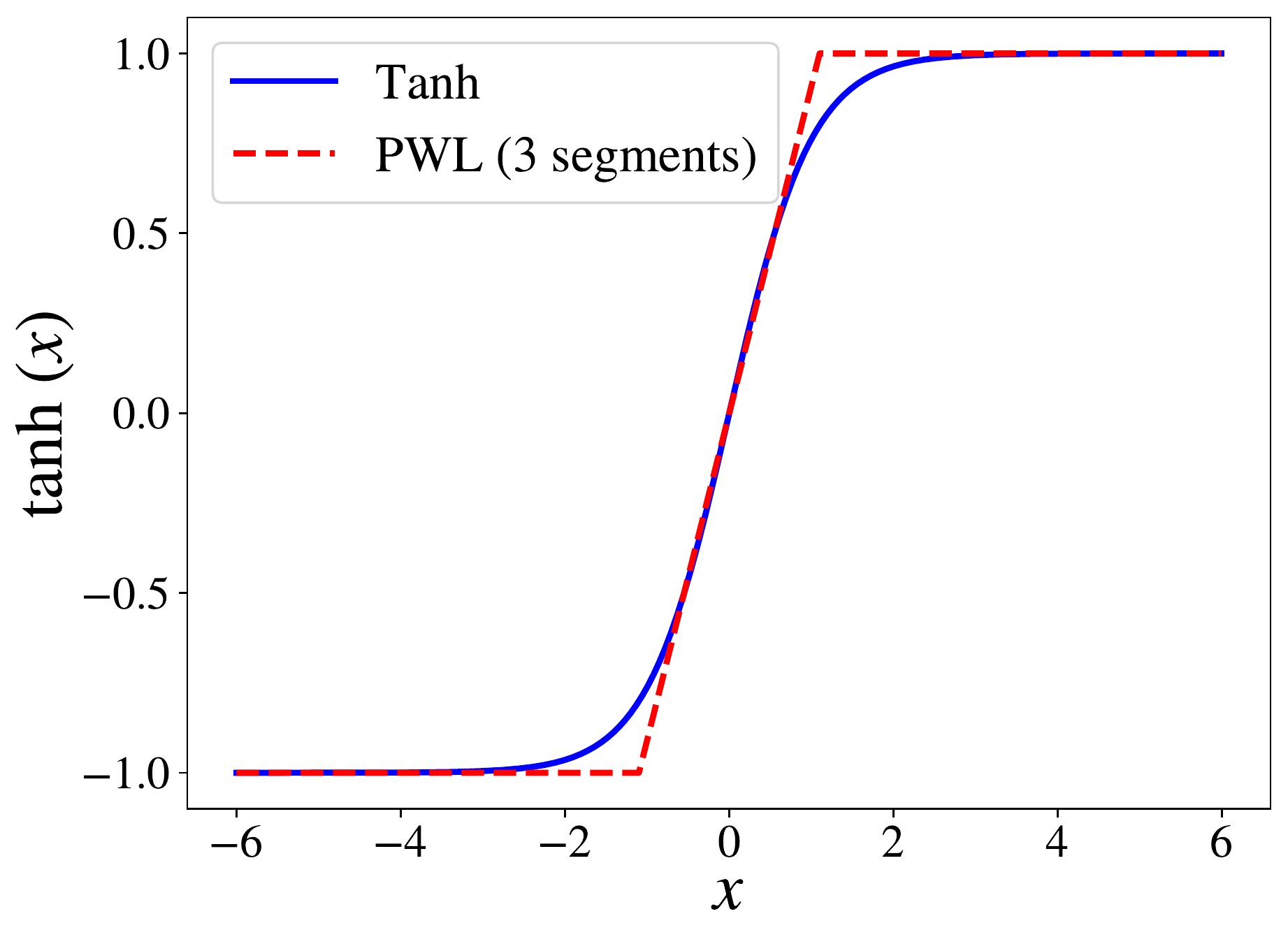}}\quad
\subcaptionbox{Tanh with 9 segments.}{\includegraphics[width=.47\linewidth]{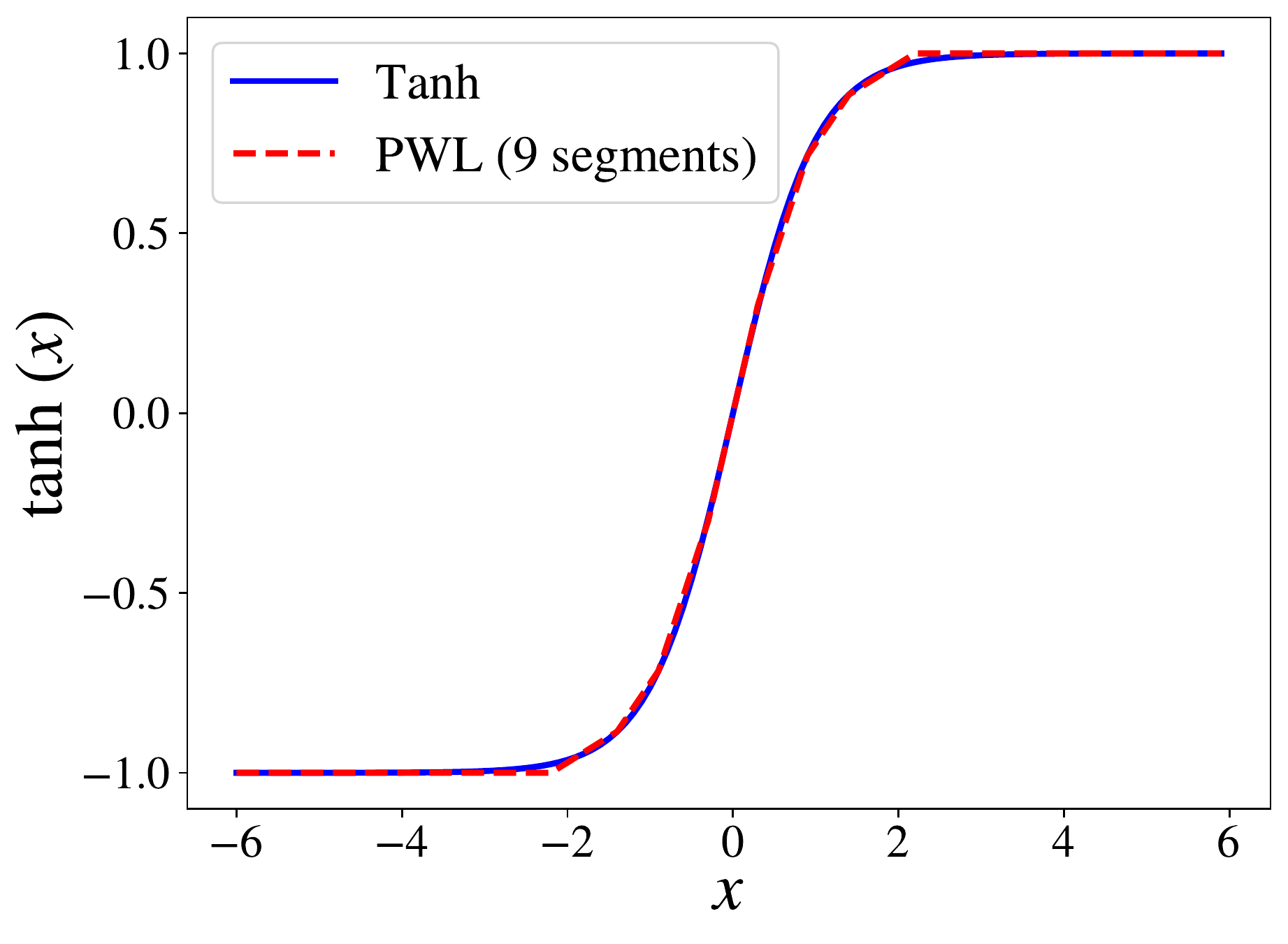}}
\medskip
\vspace{3mm}
\subcaptionbox{Sigmoid with 3 segments.}{\includegraphics[width=.47\linewidth]{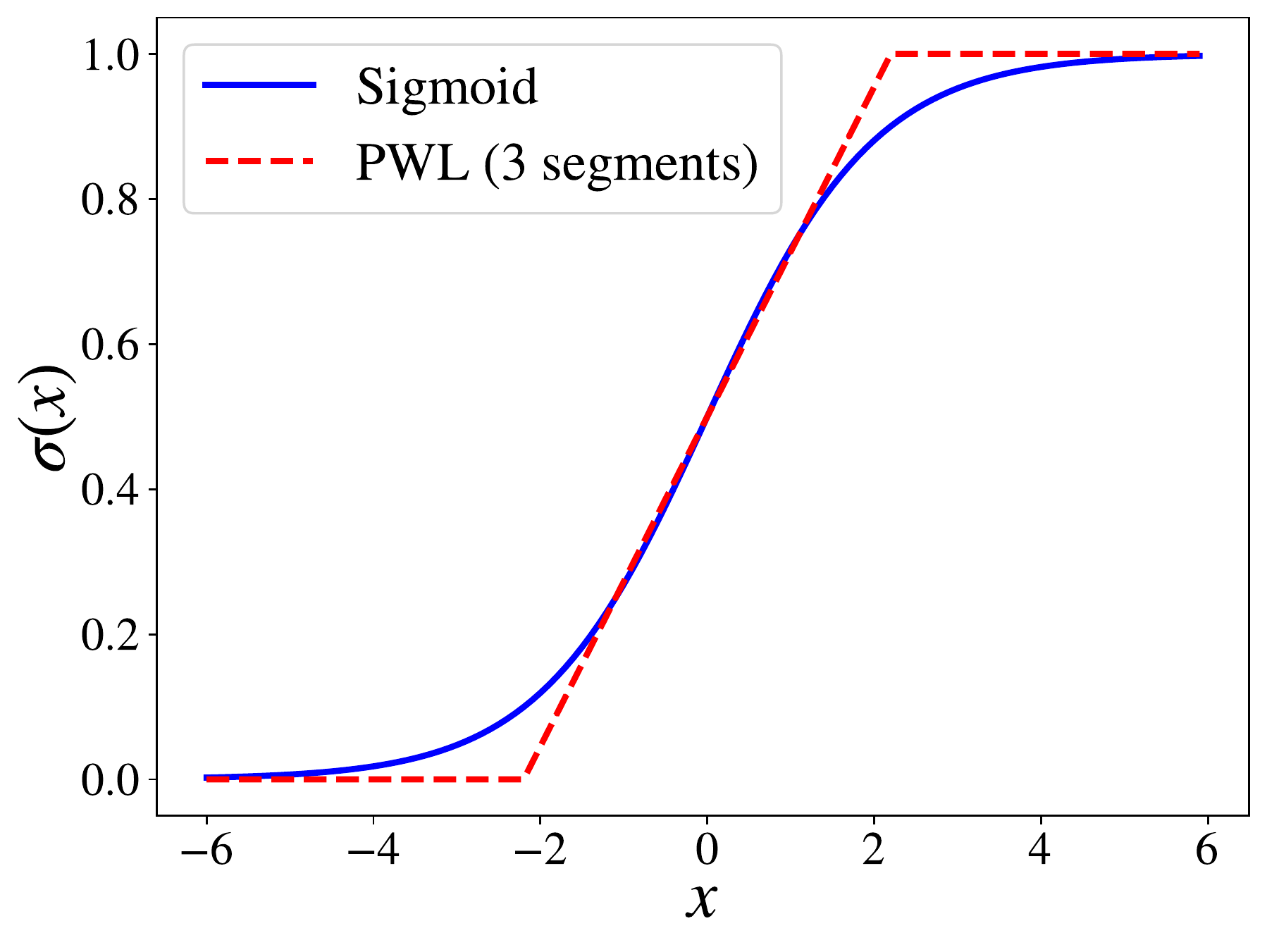}}\quad
\subcaptionbox{Sigmoid with 9 segments.}{\includegraphics[width=.47\linewidth]{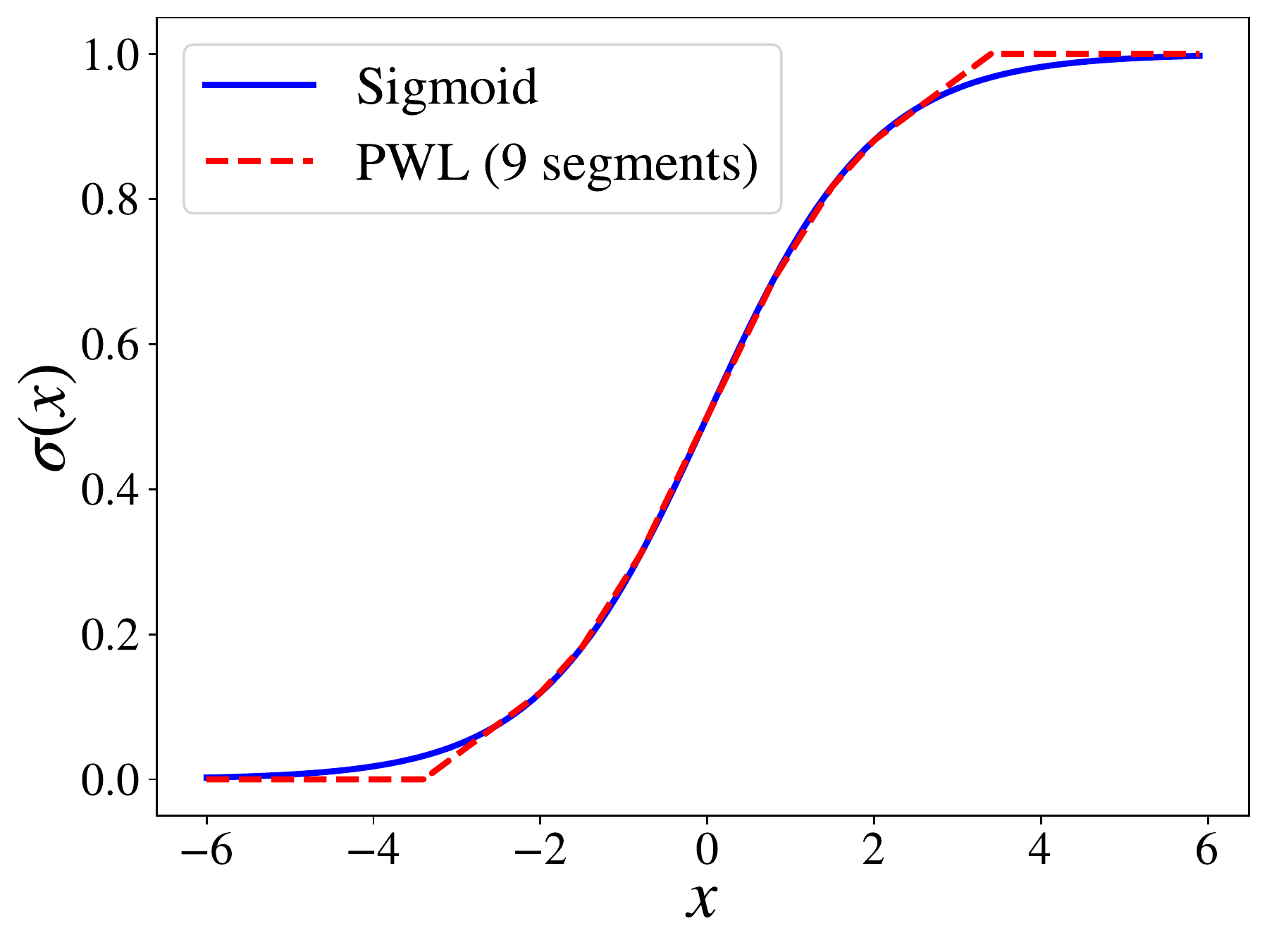}}
\caption{PWL approximation of tanh (a) -- (b) and sigmoid functions (c) -- (d).}
\label{fig:PWL}
\end{minipage}
\end{figure*}

In this section, we address the implementation of NNs' nonlinear activation function, one of the crucial components in the design of NN in hardware. In contrast to the hardware realization of the NN's weights and inputs, where we can readily proceed from the float to fixed-point representation, the activation functions' realization in hardware is not straightforward.
In an LSTM cell, the sigmoid and tanh functions are deployed as activation functions, and they are computationally expensive. Both functions contain exponential functions, making it difficult to implement them on resource-constrained hardware and requiring a large chip area \cite{ccetin2015application}. Therefore, function approximation techniques are required in place of the exact functions to realize them in the FPGA, and to reduce the overall computational complexity \cite{li2022fpga,ccetin2015application,timmons2020approximating, temurtas2004study}. 
In this work, we focus on approximating the sigmoid and tanh. We consider three different methods for the approximation: Taylor series expansion, PWL, and LUT. As shown in Fig.~\ref{fig:acti_func_fpga}, to implement the approximated activation functions on the FPGA, the FF\footnote{FF is a basic digital storage element in an FPGA, used to store the value of a digital signal and can be used in conjunction with LUTs to implement sequential logic, such as state machines and counters.}, LUT\footnote{LUT is a basic building block of an FPGA used to implement equations built from Boolean logic functions, such as AND, OR, and XOR, or to store pre-calculated values for use in arithmetic or other operations.}, and DSP slices are used to build the logic box\footnote{FPGA uses LUTs, FF and DSP slices together to implement the digital logic, memory, and computation required by the intended applications. LUTs, FFs, and DSPs are all programmable, meaning that the user can reprogram the FPGA's logic, memory, and computation elements to suit different applications.}, which takes the value $x$ and coefficients to return $\hat{y}$. The coefficients are stored in the memory as input. The coefficients define the Taylor and PWL approximations, while in the LUT approximation, they represent the quantization levels list. $\hat{y}$ is the output of the approximated activation functions, while $y$ represents the actual output of the float-precision activation function. The difference between $\hat{y}$ and $y$ is the approximation error.

The expression for the tanh function via exponential is:
\begin{equation}\label{eq.tanh}
\text{tanh } x = \frac{e^x - e^{-x} }{e^x + e^{-x}},
\end{equation}
while that for the sigmoid function reads as:
\begin{equation}\label{eq.sigmoid}
\sigma(x) =\frac{1}{1 + e^{-x}}.
\end{equation}

\subsection{Taylor Approximation Approach}
In the Taylor series approximation, the higher the degree of an approximating polynomial $n$, the better the approximation. The tanh Taylor series reads as:
\begin{equation}\label{eq.taylor.tanh}
\begin{split}
\text{tanh } x = {}& \sum_{n=0}^{\infty} \frac{2^{2n}(2^{2n} - 1) B_{2n} }{(2n)!}x^{2n-1},\;\; \text{where}{\;\lvert x \rvert < \frac{\pi}{2}}\\
= {}& x - \frac{x^3}{3} + \frac{2x^5}{15} -\frac{17x^7}{315} + \frac{62x^9}{2835} - \ldots \, ,
\end{split}
\end{equation}
where $B_{2n}$ denotes the Bernoulli number\cite{Spiegel_math_book}, $-a_t <x< a_t$, and $a_t$ is the boundary of the approximation region: when $x$ is not within $[-a_t,a_t]$, the approximation error is essential. Therefore, it is important to choose the value of $a_t$ that maximizes performance. Empirically, the slight difference in the value of $a_t$ can noticeably affect the performance. When $x$ is outside the Taylor series approximation region, we set the value of tanh $x$ to -1 or 1, according to the following expression:
\begin{equation}\label{eq.taylor.tanh.func}
  \text{tanh } \! x = \!  \begin{cases}
  1, & \mbox{ if $ x>a_t $,}\\
  x - \frac{x^3}{3} + \frac{2x^5}{15} -\frac{17x^7}{315} + \frac{62x^9}{2835}, & \mbox{ if $ -a_t <x< a_t$,}\\
  -1, & \mbox{ if $ x< -a_t $.}
  \end{cases}
\end{equation}
The plots for the different order Taylor approximations are given in Fig.~\ref{fig:taylor3_tanh}--\ref{fig:taylor9_tanh}. The value of $a_t$ is the result of the grid search, which maximizes the performance of our NN-based equalizer without re-training.

The Taylor series for the sigmoid function is:
\begin{equation}\label{eq.taylor.sigmoid}
\begin{split}
\sigma(x) ={}&\frac{1}{2} + \frac{1}{2}\text{tanh } \frac{x}{2} \\
= {}& \frac{1}{2} + \frac{x}{4} - \frac{x^3}{48} + \frac{x^5}{480} -\frac{17x^7}{80640} + \frac{31x^9}{1451520} - \ldots \, ,
\end{split}
\end{equation}
where $-a_\sigma <x< a_\sigma$ and $a_\sigma$ is the point where the Taylor series approximation of the sigmoid starts to diverge. Similarly to tanh, the values of the sigmoid approximation in regions less than $-a_\sigma$ and greater than $a_\sigma$ are set to 0 and 1, respectively, as follows:
\begin{equation}\label{eq.taylor.sigmoid.func}
  \sigma(x)\!=\!\begin{cases}
  1, \!\!\! &\! \mbox{if $ x>a_\sigma $,}\\
  \frac{1}{2}\! +\! \frac{x}{4}\! -\! \frac{x^3}{48}\! +\! \frac{x^5}{480}\! -\!\frac{17x^7}{80640}\! +\! \frac{31x^9}{1451520},\!\!\! &\! \mbox{if $-a_\sigma\! <\!x\!<\! a_\sigma$,}\\
  0,\!\!\! &\! \mbox{if $x< -a_\sigma $.}
  \end{cases}
\end{equation}
 The Taylor approximation plots corresponding to Eq.~(\ref{eq.taylor.sigmoid.func}), when the highest order of the polynomial is 3 and 9, are given in  Fig.~\ref{fig:taylor3_sigmoid}--\ref{fig:taylor9_sigmoid}.

We evaluate the performance (in terms of Q-factor) when the approximation for both tanh and sigmoid functions is carried out simultaneously, with different orders of the approximating polynomial up to 9\textsuperscript{th} order. The values of $a_t$ and $a_\sigma$ are chosen by using the grid search, aiming to maximize the Q-factor when replacing the exact activation functions with their Taylor series approximation without re-training the weights. 
The Taylor series approximation reduces the computational cost and time required to compute the activation function considerably, compared to the processing using the original function\cite{timmons2020approximating}. 

\subsection{Piecewise Linear Approximation Approach}
The PWL approximation, introduced in~\cite{amin1997piecewise}, is a combination of linear segments that approximates the activation or nonlinear function \cite{basterretxea2004approximation, li2022fpga}. Increasing the number of linear segments to represent the nonlinear function allows us to achieve better accuracy. The PWL approximation is a promising method to reach a higher processing speed since it consumes fewer resources on FPGA\footnote{\cite{tommiska2003efficient} shows that the implementation of PWL can be further optimized to have zero multipliers by simplifying the shift and addition operations.} compared to the Taylor approximation: to reach higher accuracy, the Taylor approach fits the nonlinear function with high-order expressions, which results in the consumption of resources, while the PWL can reach the same level of accuracy with the use of more segments, but without employing high-order operations\cite{li2022fpga}.

In this work, we compare the performance of our NN-based equalizers when applying 3-, 5-, 7-, and 9-segment  PWL approximations to both tanh and sigmoid\footnote{Note that when the number of segments is lower than 3 segments that used to represent sigmoid or tanh in the biLSTM cell, the biLSTM model in our case is not able to learn to mitigate the approximation errors.}. The expressions for the PWL used in this paper are included in Table~\ref{tab:piecewise} in Appendix \ref{app:act_func}. The corresponding plots for the equations mentioned in Table~\ref{tab:piecewise} with 3 and 9 segments are depicted in Fig.~\ref{fig:PWL}a--\ref{fig:PWL}b for tanh, and Fig.~\ref{fig:PWL}c--\ref{fig:PWL}d for sigmoid. Note that we use grid search to find the coefficients for each expression, aiming to maximize the performance in terms of Q-factor, after the actual activation functions are replaced by the approximations over the trained weights, instead of minimizing the difference/areas between the exact function and the approximation curves. It is carried out because, in our case, minimizing the difference/areas between the curves noticeably degrades the Q-factor performance of the NN equalizer when the NN predicts the output with the replaced approximated activation functions.

\subsection{Lookup Table Approximation Approach}

\begin{figure}[b]
    \centering
    \includegraphics[width=0.7\linewidth,scale = 0.3]{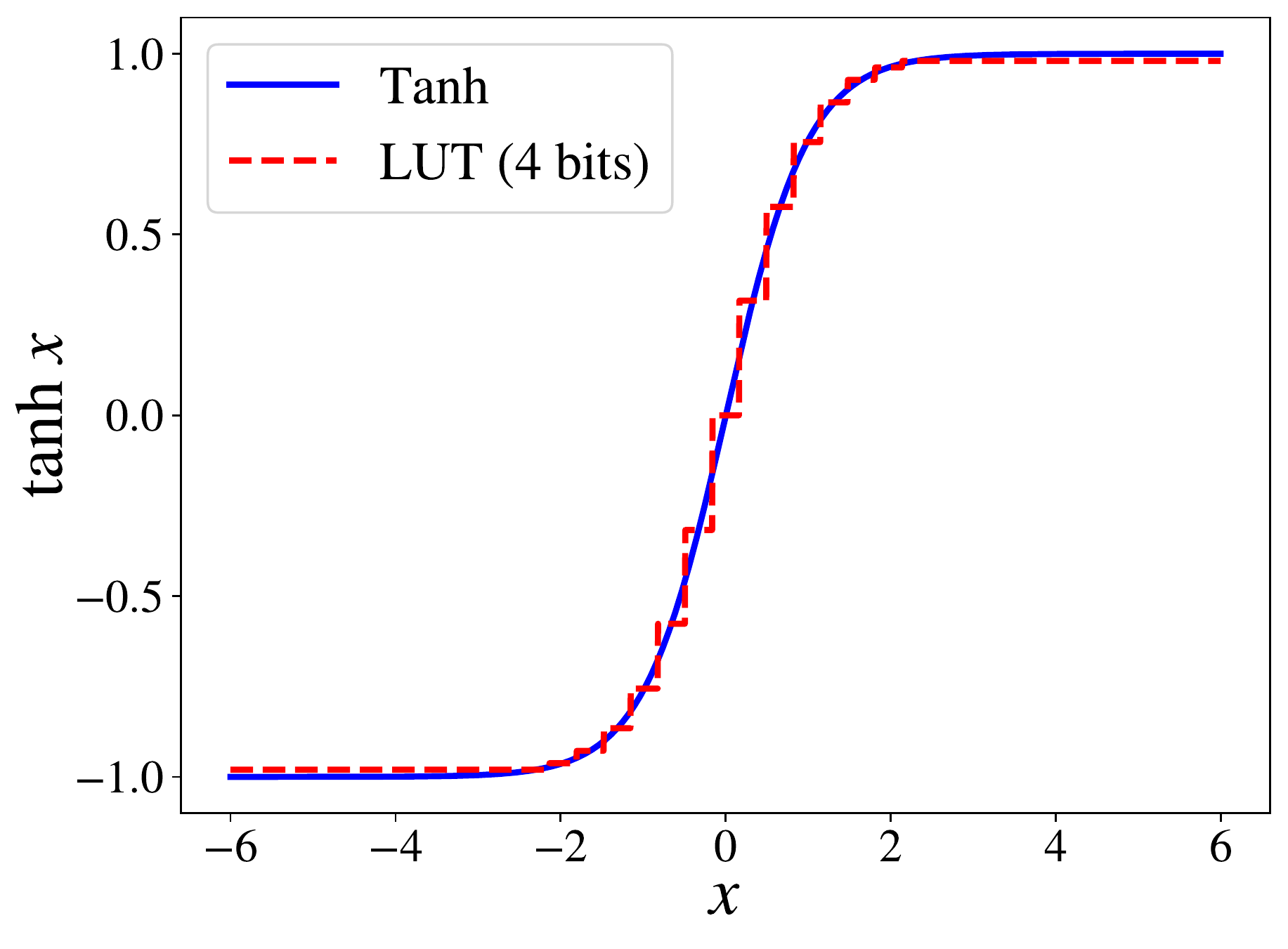}
    \caption{LUT approximation of tanh function with the number of bits equal to 4.}
    \label{fig:lut_tanh}
\end{figure}

\begin{figure*}[th]
  \centering
\begin{subfigure}{.32\textwidth}
  \centering
    \includegraphics[scale = 0.3]{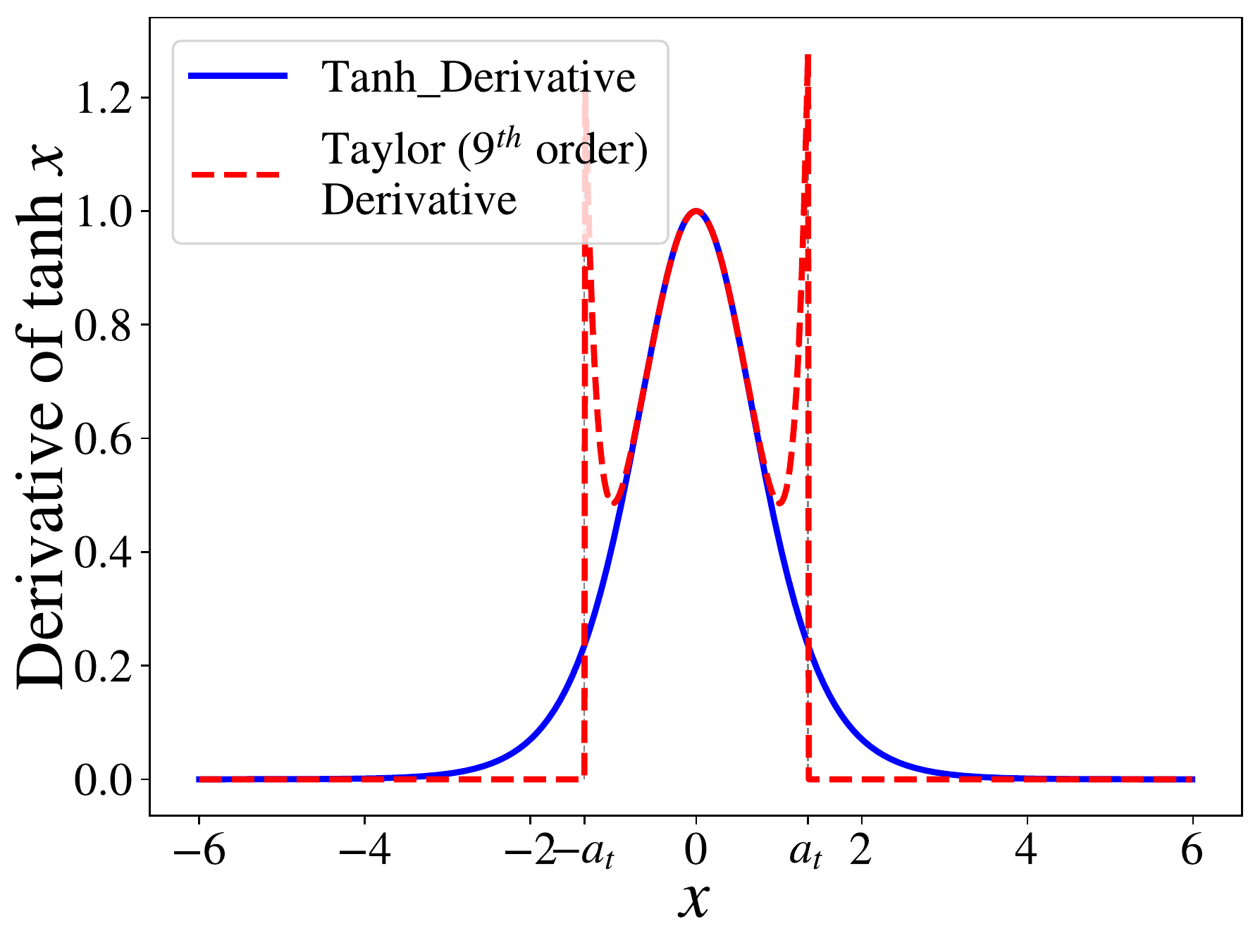}
\caption{Gradient of Taylor approximation.}
\label{fig:derivative_taylor}
\end{subfigure}
\hfill
\begin{subfigure}{.32\textwidth}
  \centering
  \includegraphics[scale = 0.3]{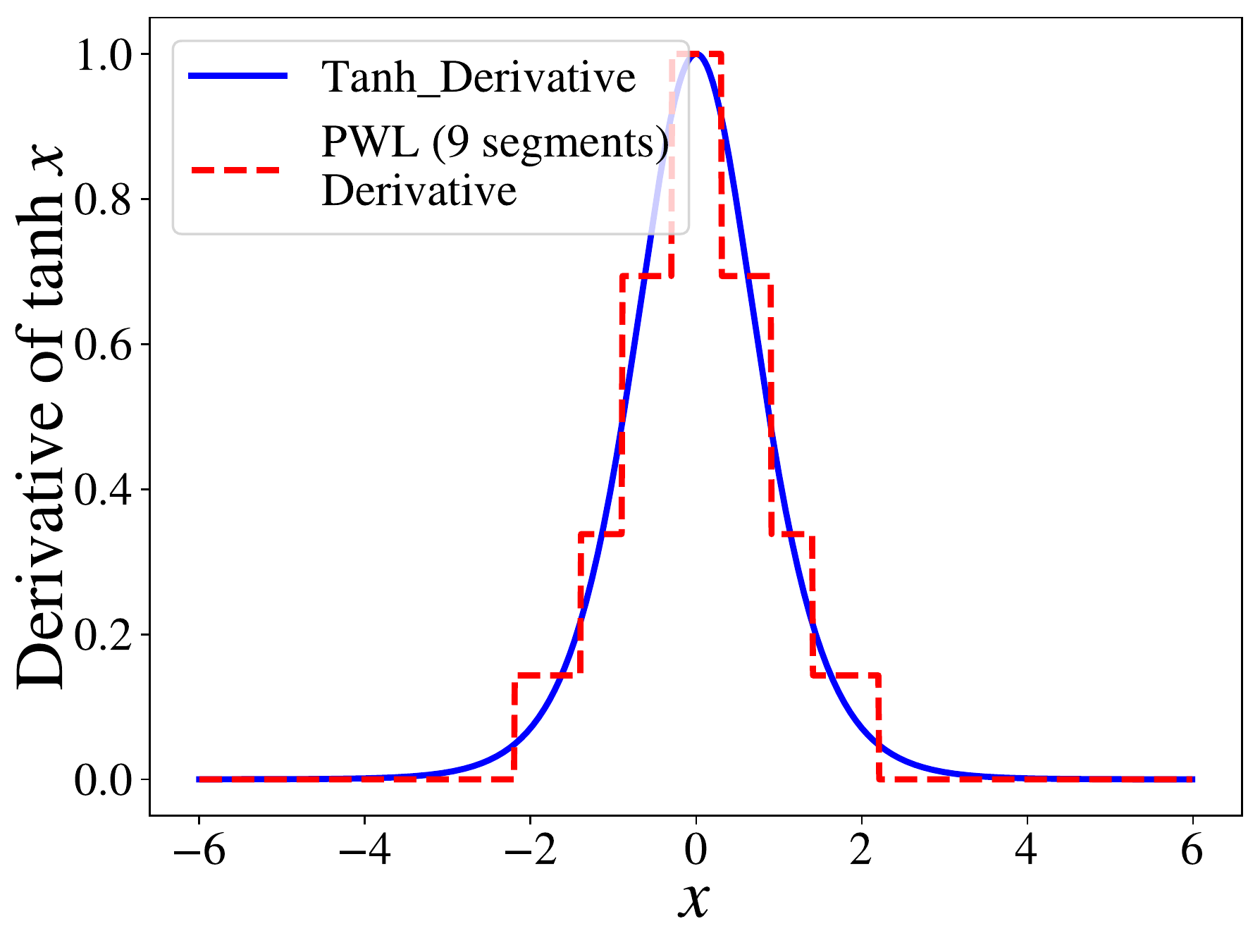}
\caption{Gradient of PWL approximation.}
\label{fig:derivative_PWL}
\end{subfigure}
\hfill
\begin{subfigure}{.32\textwidth}
  \centering  
  \includegraphics[scale = 0.3]{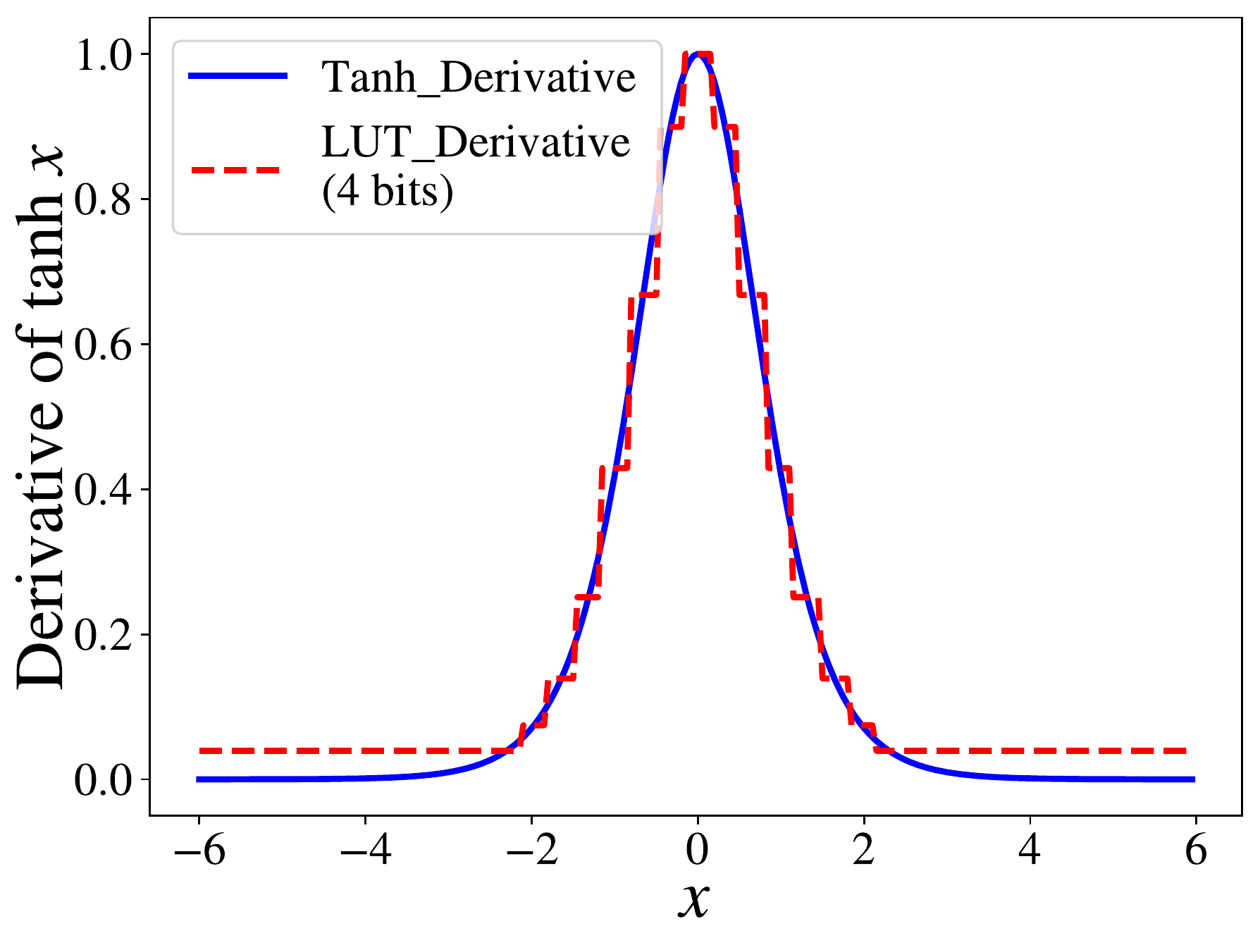}
\caption{Gradient of LUT approximation.}
\label{fig:derivative_LUT}
\end{subfigure}
\caption{The derivative of the tanh function for the approximations using (a) Taylor series with the highest order of 9, (b) PWL with 9 segments, (c) LUT approximation with the number of bits equal to 4.}
\label{fig:derivative}
\end{figure*}

The LUT approximation is a commonly used method for the activation functions' hardware implementation \cite{yang2018design}. The LUT approximates the function with a limited number of uniformly distributed points. This approach offers a high-performance design, and the fastest implementation compared to other methods. At the same time, a large amount of memory is required to store the LUT on the hardware \cite{namin2009artificial, namin2009efficient}. The chip area requirements for the LUT approximation grow exponentially with the required approximation accuracy \cite{namin2009efficient}. The number of bits used to represent values in the LUT directly affects the approximation error and the required memory size. An example of the LUT approximation of tanh with the number of bits equal to 4 is presented in Fig.~\ref{fig:lut_tanh}.

The LUT approach is similar to traditional quantization, in which full precision values are assigned to uniform quantization levels, i.e. the value $x$ is mapped to $\hat{x}$ which is the closest value of $x$ in the quantization level list \cite{wang2022learnable}. The LUT stores the values of the quantization levels ($\hat{x}$) and their corresponding $f(\hat{x})$, in our case $tanh(\hat{x})$ or $\sigma(\hat{x})$. The difference between the exact value $f(x)$ (the blue curve in Fig.~\ref{fig:lut_tanh}) and the approximation $f(\hat{x})$ (the red curve in Fig.~\ref{fig:lut_tanh}) introduces the approximation errors.

We investigate the Q-factor performance of our model for the LUT representation of activation functions when the number of bits used ranges from 2 to 16. 

\subsection{Reducing Approximation Error through the Learning via Stochastic Gradient Descent}
\label{sec:re_training_act}
Once the activation functions are replaced by the approximation, the NN performance can drastically drop. However, training the model with approximated activation functions can enhance the performance because the model learns to reduce the approximation error. Stochastic gradient descent (SGD) is the training approach that we apply in this work. 
The training can be undertaken from scratch, which means that the NN is trained when the activation functions are replaced by approximations from the beginning without any pre-assigned weights. Another approach to training is to use the weights of the model pre-trained with the true activation functions, then re-train the model after the replacement of the approximations to learn the approximation errors. The latter results in a considerably shorter training time. We report the results of the second method because, in our tests, training from scratch takes significantly longer to converge and sometimes can provide even worse results. It is worth noting that another available training approach is to only train the coefficients of the Taylor and PWL equations without re-training the NN weights; however, in our case, the performance was not acceptable when using a low number of segments in PWL and training with this approach.

To train the NN with the approximation of the activation function via the SGD, the gradient of the approximation function must be computed. For the Taylor approximation, the Taylor series gradient is calculated with respect to the Taylor series approximation equations Eq.~(\ref{eq.taylor.tanh.func}) for tanh and Eq.~(\ref{eq.taylor.sigmoid}). Fig.~\ref{fig:derivative_taylor} shows an example of the derivative of the tanh approximation using the Taylor series with the highest order of 9; the gradient (red curve) is not smooth due to the polynomial nature of the Taylor series. This fact can limit the training ability, especially when training from scratch, as noted in Section~\ref{sec:result_act}. Concerning the PWL, the gradient is the slope of the expressions from Table~\ref{tab:piecewise} (in the Appendix section). Fig.~\ref{fig:derivative_PWL} depicts the gradient of the PWL approximation with 9 segments. Note that due to the non-differentiability of LUT, it is challenging to learn the LUT-approximated model \cite{wang2022learnable}. In this work, to train the LUT, we generate LUTs for the gradient of both sigmoid and tanh for each interval of the LUT approximations. Fig.~\ref{fig:derivative_LUT} shows the gradient of the tanh LUT with 4 bits, corresponding to the tanh approximation in Fig.~\ref{fig:lut_tanh}.


\section{Results and Discussions} \label{sec:results}
\subsection{Experimental and Numerical Setups} \label{sebsec:setup}
\begin{figure}[th!]
\centering
\includegraphics[width=0.45\textwidth]{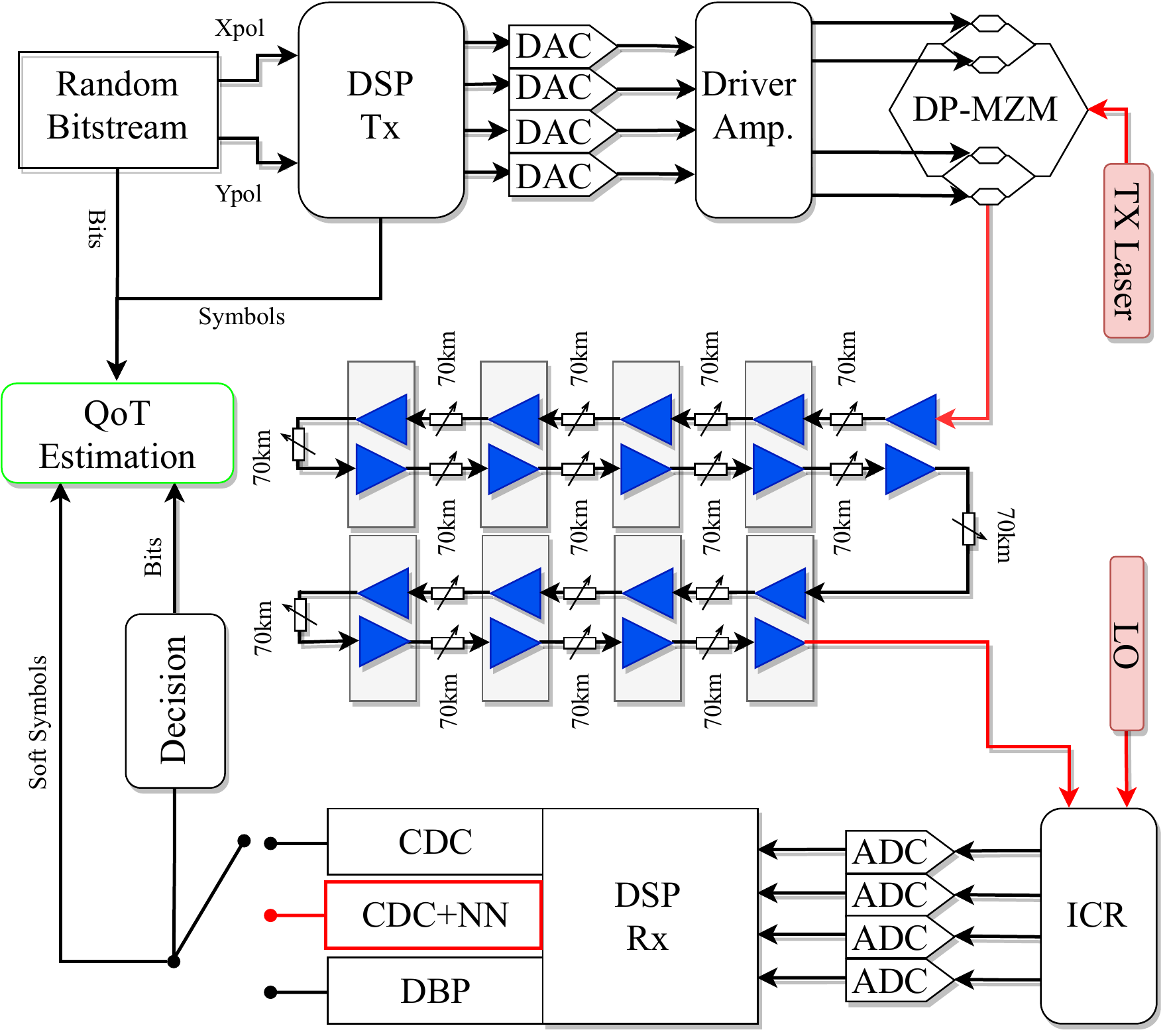}
\caption{Experimental setup. The input of the NN (shown as the red rectangle after DSP RX) is the soft output of the regular DSP before the decision unit.}
\label{setup}
\end{figure}

We assess the performance of the NN-based equalizers with reduced complexity by using the data not only from numerical simulations but also from a real experimental setup, to make our analysis as complete as possible. The setup used in our experiment is shown in Fig.~\ref{setup}. 
At the transmitter, a DP-16QAM 34~GBd symbol sequence was mapped out of the data bits generated by a Mersenne Twister algorithm \cite{matsumoto1998mersenne}. Then, a digital RRC filter with 0.1 roll-off was applied. The resulting filtered digital samples were resampled and uploaded to a digital-to-analog converter (DAC) operating at 88 GSamples/s. The output of the DAC was amplified by a four-channel electrical amplifier that drove a dual-polarization IQ Mach-Zehnder modulator, modulating the continuous waveform carrier produced by an external
cavity laser at the wavelength $\lambda = 1.55 \, \rm{\mu m}$. 
The resulting optical signal was transmitted over 17$\times$70 km spans of LEAF. Erbium-doped fiber amplifiers (EDFAs) are used to compensate for the loss in each fiber span at their output. The EDFA's noise figure was in a 4.5 to 5 dB range. The parameters of the LEAF are: the attenuation coefficient $ \alpha = 0.225$~dB/km, the chromatic dispersion coefficient $D = 4.2$~ps/(nm$\cdot$km), and the effective nonlinear coefficient $\gamma$ = 2~(W$\cdot$ km)$^{-1}$.

\begin{figure*}[t]
\centering
\begin{subfigure}{.45\textwidth}
  \centering
         \begin{tikzpicture}[scale=0.8]
    \begin{axis} [ylabel={BER}, 
        xlabel={Launch power [dBm]},
        ylabel={Q-Factor [dB]},
        grid=both,  
         ylabel near ticks,
        xmin=-4, xmax=4,
    	xtick={-4, ..., 4},
    	ymin=0, ymax=6,
        legend style={legend pos=south west, legend cell align=left,fill=white, fill opacity=0.6, draw opacity=1,text opacity=1},
    	grid style={dashed}]
        ]
        \addplot[color=blue, mark=square, very thick]     coordinates {
    (-4, 3.12)(-3, 3.62)(-2, 4.06)(-1, 4.41)(0, 4.64)(1,4.68)(2,4.48)(3,3.96)(4,3.1)
    };
    \addlegendentry{Deep CNN eq.};
    
    \addplot[color=red, mark=*, very thick]   
    coordinates {
    (-4, 3.2)(-3, 3.71)(-2, 4.18)(-1, 4.56)(0, 4.95)(1,5.18)(2,5.24)(3,5.06)(4,4.51)
    };
    \addlegendentry{biLSTM eq.};

        \addplot[color=green, mark=+,very thick]    coordinates {
    (-4, 3.16)(-3, 3.69)(-2, 4.19)(-1, 4.63)(0, 4.98)(1,5.19)(2,5.21)(3,4.98)(4,4.43)
    };
    \addlegendentry{DBP 1 StpS};
        \addplot[color=black,mark=triangle, very thick]    coordinates {
    (-4,3.01)(-3, 3.42)(-2, 3.74)(-1, 3.91)(0, 3.84)(1,3.52)(2,2.87)(3,1.94)(4,0.72)
    };
    \addlegendentry{CDC (Regular DSP)};
        \end{axis}
       
    \draw[red!80!pink,dashed] (2.6,3.7) -- (5.8,3.7);
    \draw[red!80!pink,dashed] (2.6,5.) -- (5.8,5.);
    \draw[thick, <->,red] (5,3.7) -- +(0,1.3);
    \node[text width=1cm] at (4.3,4.1) 
    {\textcolor{red}{ 1.3dB}};

    \end{tikzpicture}
\caption{Simulation.}  
\label{fig:performance_result_sim}
\end{subfigure}
\hfill
\begin{subfigure}{.45\textwidth}
    \centering
         \begin{tikzpicture}[scale=0.8]
    \begin{axis} [ylabel={BER}, 
        xlabel={Launch power [dBm]},
        ylabel={Q-Factor [dB]},
        grid=both,  
         ylabel near ticks,
        xmin=-4, xmax=4,
    	xtick={-4, ..., 4},
    	ymin=0, ymax=6,
        legend style={legend pos=south west, legend cell align=left,fill=white, fill opacity=0.6, draw opacity=1,text opacity=1},
    	grid style={dashed}]
        ]
        \addplot[color=blue, mark=square, very thick]     coordinates {
    (-4, 3.15)(-3, 3.57)(-2, 4.05)(-1, 4.5)(0, 4.82)(1,4.9)(2,4.5)(3,3.88)(4,2.64)
    };
    \addlegendentry{Deep CNN eq.};
    
    \addplot[color=red, mark=*, very thick]   
    coordinates {
    (-4, 3.48)(-3, 4)(-2, 4.52)(-1, 5.1)(0, 5.54)(1,5.66)(2,5.54)(3,5.1)(4,4.2)
    };
    \addlegendentry{biLSTM eq.};

        \addplot[color=green, mark=+,very thick]    coordinates {
    (-4, 3.22)(-3, 3.72)(-2, 4.16)(-1, 4.59)(0, 5.04)(1,5.47)(2,5.41)(3,5.1)(4,4.11)
    };
    \addlegendentry{DBP 1 StpS};
        \addplot[color=black,mark=triangle, very thick]    coordinates {
    (-4,3)(-3, 3.4)(-2, 3.7)(-1, 3.92)(0, 3.94)(1,3.5)(2,2.7)(3,1.61)(4,0.1)
    };
    \addlegendentry{CDC (Regular DSP)};
        \end{axis}
    \draw[red!80!pink,dashed] (1.8,3.75) -- (4.8,3.75);
    \draw[red!80!pink,dashed] (1.8,5.4) -- (4.8,5.4);
    \draw[thick, <->,red] (3.8,3.75) -- +(0,1.6);
        \node[text width=1cm] at (4.55,4.1) 
    {\textcolor{red}{1.7dB}};

    \end{tikzpicture}
\caption{Experiment.}
\label{fig:performance_result_exp}
\end{subfigure}
\caption{Q-factor versus launch power for (a) simulation and (b) experiment corresponding to the transmission of an SC-DP 16QAM 34 GBd signal along 17$\times$70km of LEAF. The difference between the  time domain CDC and the biLSTM equalizer's results is marked with red arrows. The case of the floating-point models' accuracy for the different types of NN equalizers (described in the legends), together with the 1StpS DBP and CDC performance curves.}
\label{fig:performance_result}
\end{figure*}
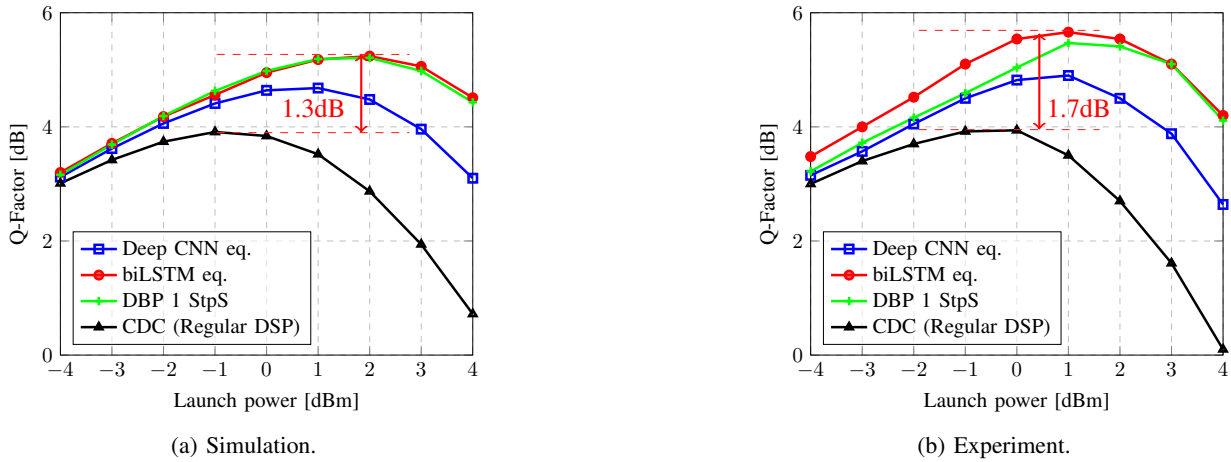

On the Rx side, the optical signal was converted to the electrical domain using an integrated coherent receiver. The resulting signal was sampled at 80~Gsamples/s with a digital sampling oscilloscope and processed by an offline DSP based on the algorithms described in~\cite{kuschnerov2010data}. First, the bulk accumulated dispersion was compensated using a frequency domain equalizer, which was followed by the mitigation of the carrier frequency offset. A constant-amplitude zero autocorrelation (CAZAC)-based training sequence was then located in the received frame, and the equalizer transfer function was estimated from it. Afterward, the two polarizations were demultiplexed, and the signal was corrected for clock frequency and phase offsets. The carrier phase estimation was then carried out with the help of pilot symbols. Subsequently, the resulting soft symbols were used as the input for the NN equalizer. Finally, the pre-FEC BER was evaluated from the signal at the NN output. The performance of the system was evaluated in terms of the Q-factor, expressed through the BER as $Q = 20 \: \mathrm{log_{10}} \left[\sqrt{2} \: \mathrm{erfc^{-1}}(2\,\rm BER)\right]$.

Concerning the simulation, we tried to mimic the experimental transmission scenario. The propagation of the signal along the fiber was simulated by solving the Manakov equations using the split-step Fourier method with a step size of 1 km. At the receiver, after the full CDC (time domain) and downsampling to the symbol rate, the received symbols were normalized to the transmitted ones. The normalization process can be viewed as its normalization by a constant $K_\text{DSP}$ learned using the following equation:
\begin{equation}
    \label{eqn:norming}
    \mathcal{K}_\text{DSP} = \min_\mathcal{K}\left\|\mathcal{K}\cdot x_{h\!/\!v}(z,t) - x_{h\!/\!v}(0,t)\right\|,
\end{equation}
where the constants $\mathcal{K}, \, \mathcal{K}_\text{DSP} \in \mathbb{C}$ and $x_{h\!/\!v}$ is the signal in $h$ or $v$ polarization. Furthermore, the Gaussian noise was added to the data signal, as to represent the additional transceiver components-induced distortions present in the experiment. As a result, the Q-factor level of the simulated data (without NN equalization) was matched to the experimental one. Note that the polarization mode dispersion (PMD) is not considered in the simulation. In the experimental data, PMD was already compensated by Infinera's DSP, so we can say that our study is not influenced by PMD.

Finally, unlike the NN equalizer, which operates with 1 Sa/symbol, the DBP used to benchmark the performance curves (the implementation described in~\cite{napoli2014reduced}), operated with 2.3 Sa/symbol (and with 1~StpS with the scheme parameters optimized for the best performance). Regarding the CDC implementation, we designed a time-domain equalizer as in~\cite{xu2010chromatic} with 517 taps in C++. For the realization in hardware, we followed the same design steps 3 and 4 described in Sec.~\ref{sec:pipeline} for the NN implementation. To be more specific, the goal of this result section is to assess the complexity of NN with respect to CDC, while guaranteeing a level of nonlinear compensation comparable to one of the widely used DBP. The CDC benchmark is the most important because our primary goal is to show the readiness of NN with respect to the already available algorithm in commercial transponders. In contrast, none of the existing DBP versions has reached the hardware level of implementation. In this context, this paper shows that the NN-based equalizer achieves a performance similar to that obtained with the DBP~\cite{napoli2014reduced} while approaching the complexity of the CDC block.

\begin{figure*}[bh]
  \centering
\begin{subfigure}{.30\textwidth}
  \centering
\begin{tikzpicture}[scale=0.7]
    \begin{axis} [
        xlabel={Highest polynomial order},
        ylabel={Q-Factor [dB]},
        grid=both,  
        ylabel near ticks,
        xmin=3, xmax=9,
    	xtick={3,5,7,9},
    	ymin=0, ymax=5.5,
        legend style={legend pos=south east, legend cell align=left,fill=white, fill opacity=0.6, draw opacity=1,text opacity=1},
    	grid style={dashed}]
        ]
    \addplot[color=green, very thick]     coordinates {
    (2, 5.2)(16, 5.2)
    };
    \addlegendentry{Without approximation};
    
    \addplot[color=red, mark=*, dashed, very thick]     coordinates {
    (3, 1.19)(5, 3.08)(7, 3.5)(9, 4.18)
    };
    \addlegendentry{Without re-training};
    
    \addplot[color=red, mark=*, very thick]     coordinates {
    (3, 5.08)(5, 5.13)(7, 5.14)(9, 5.13)
    };
    \addlegendentry{With re-training};
 
        \end{axis}
\end{tikzpicture}     
\caption{Taylor series approximation.}
\label{fig:performance_vs_comp_taylor}
\end{subfigure}
\hfill
\begin{subfigure}{.30\textwidth}
  \centering
\begin{tikzpicture}[scale=0.7]
    \begin{axis} [
        xlabel={No. of segments},
        ylabel={Q-Factor [dB]},
        grid=both,  
        ylabel near ticks,
        xmin=3, xmax=9,
    	xtick={3,5,7,9},
    	ymin=0, ymax=5.5,
        legend style={legend pos=south east, legend cell align=left,fill=white, fill opacity=0.6, draw opacity=1,text opacity=1},
    	grid style={dashed}]
        ]
        \addplot[color=green, very thick]     coordinates {
    (2, 5.2)(16, 5.2)
    };
    \addlegendentry{Without approximation};
        \addplot[color=blue, mark=*, dashed, very thick]     coordinates {
    (3, 0)(5, 3.21)(7, 4.26)(9, 4.48)
    };
    \addlegendentry{Without re-training};
    
    \addplot[color=blue, mark=*, very thick]     coordinates {
    (3, 5.09)(5, 5.075)(7, 5.1)(9, 5.1)
    };
    \addlegendentry{With re-training};
 
        \end{axis}
\end{tikzpicture}     
\caption{PWL approximation.}
\label{fig:performance_vs_comp_PWL}
\end{subfigure}
\hfill
\begin{subfigure}{.30\textwidth}
  \centering
\begin{tikzpicture}[scale=0.7]
    \begin{axis} [
        xlabel={No. of bits},
        ylabel={Q-Factor [dB]},
        grid=both,  
        ylabel near ticks,
        xmin=2, xmax=16,
    	ymin=0, ymax=5.5,
        legend style={legend pos=south east, legend cell align=left,fill=white, fill opacity=0.6, draw opacity=1,text opacity=1},
    	grid style={dashed}]
        ]
    \addplot[color=green, very thick]     coordinates {
    (2, 5.2)(16, 5.2)
    };
    \addlegendentry{Without approximation};
    
    \addplot[color=violet, mark=*, dashed, very thick]     coordinates {
    (2, 0)(3, 0)(4, 0)(5, 0)(6, 0)(7, 2.74)(8, 4.44)(9, 5.00)(10, 5.16)(11, 5.188)(12, 5.197)(13, 5.199)(14, 5.2)(15, 5.196)(16, 5.197)
    };
    \addlegendentry{Without re-training};
    
    \addplot[color=violet, mark=*, very thick]     coordinates {
    (2, 0)(3, 0.042)(4, 2.45)(5, 3.03)(6,3.45)(7, 4.53)(8, 4.85)(9, 5)(10, 5.15)(11, 5.18)(12, 5.2)(13, 5.192)(14, 5.21)(15, 5.196)(16, 5.197) 
    };
    \addlegendentry{With re-training};
 
        \end{axis}
\end{tikzpicture}     
\caption{LUT approximation.}
\label{fig:performance_vs_comp_LUT}
\end{subfigure}
\caption{Q-factor versus complexity in terms of polynomial order for the Taylor approximation, pane (a), for the number of segments for PWL approximation, (b), and for the number of bits for the LUT, (c).}
\label{fig:performance_vs_comp}
\end{figure*}
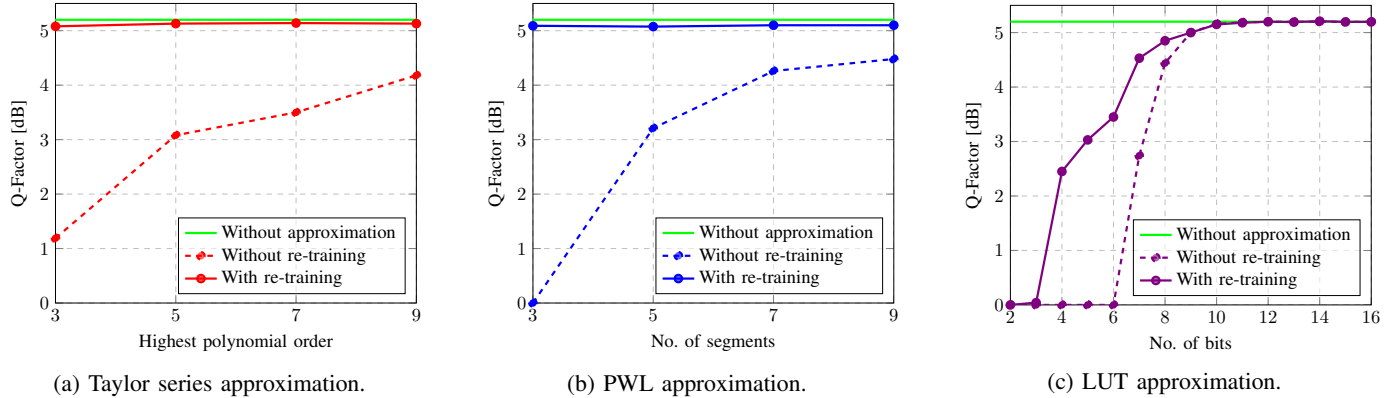

\begin{figure}[th]
    \centering
\scalebox{1.4}{\input{convergence_speed.tikz}}
\caption{Convergence study of the re-training to mitigate the approximation errors of Taylor series (3$^\text{rd}$ order), PWL (3 segments), and LUT ($n_\text{bit}=7$) approximations.}
\label{convergency_AF}
\end{figure}
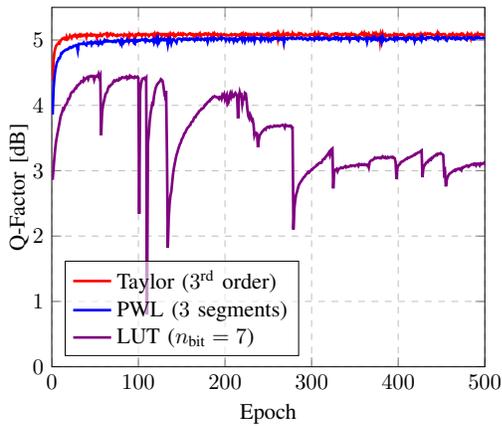

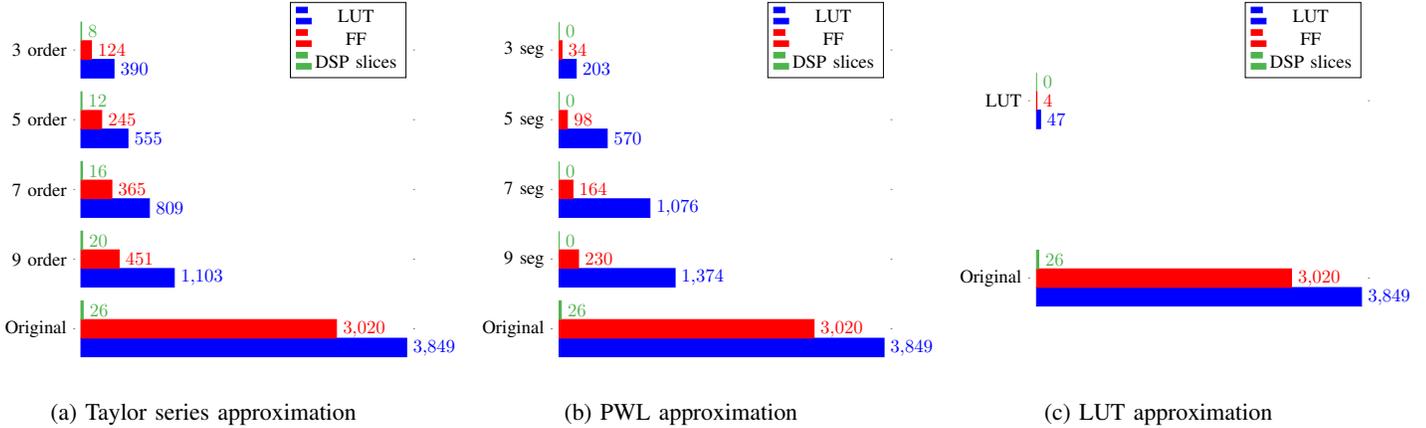
\begin{figure*}[th!]
 \centering
\begin{subfigure}{.30\textwidth}
  \centering
\begin{tikzpicture}[scale=0.7]
  \begin{axis}[
    xbar,
    y axis line style = { opacity = 0 },
    axis x line       = none,
    tickwidth         = 1pt,
    enlarge y limits=0.2,
    width=8cm,
    height=9cm,
    xbar=0.1,
    enlarge x limits  = 0.02,
    symbolic y coords = {Original, 9 order, 7 order, 5 order,3 order},
    nodes near coords,
  ]
  \addplot [style={blue,fill=blue,mark=none}] coordinates { (3849,Original)(1103,9 order)(809,7 order) (555,5 order)  (390,3 order) };
  \addplot[style={red,fill=red,mark=none}] coordinates { (3020,Original)(451,9 order)(365,7 order) (245,5 order)   (124,3 order)  };
    \addplot [style={green!40!gray,fill=green!40!gray,mark=none}]coordinates { (26,Original)(20,9 order)(16,7 order) (12,5 order)(8,3 order)};
  \legend{LUT, FF, DSP slices}
  \end{axis}
\end{tikzpicture}   
\caption{Taylor series approximation}
\label{fig:complexity_taylor}
\end{subfigure}
\hfill
\begin{subfigure}{.30\textwidth}
  \centering
\begin{tikzpicture}[scale=0.7]
  \begin{axis}[
    xbar,
    y axis line style = { opacity = 0 },
    axis x line       = none,
    tickwidth         = 1pt,
    enlarge y limits=0.2,
    width=8cm,
    height=9cm,
    xbar=0.1,
    enlarge x limits  = 0.02,
    symbolic y coords = {Original, 9 seg, 7 seg, 5 seg,3 seg},
    nodes near coords,
  ]
  \addplot [style={blue,fill=blue,mark=none}] coordinates { (3849,Original)(1374,9 seg)(1076,7 seg) (570,5 seg)(203,3 seg) };
  \addplot[style={red,fill=red,mark=none}] coordinates { (3020,Original)(230,9 seg)(164,7 seg) (98,5 seg)(34,3 seg)  };
  \addplot[style={green!40!gray,fill=green!40!gray,mark=none}]coordinates { (26,Original)(0,9 seg)(0,7 seg) (0,5 seg)   (0,3 seg)  };
  \legend{LUT, FF, DSP slices}
  \end{axis}
\end{tikzpicture}   
\caption{PWL approximation}
\label{fig:complexity_PWL}
\end{subfigure}
\hfill
\begin{subfigure}{.30\textwidth}
  \centering
\begin{tikzpicture}[scale=0.7]
  \begin{axis}[
    xbar,
    y axis line style = { opacity = 0 },
    axis x line       = none,
    tickwidth         = 1pt,
    enlarge y limits=0.6,
    width=8cm,
    height=9cm,
    xbar=0.1,
    ytick=data,
    enlarge x limits  = 0.02,
    symbolic y coords = {Original, LUT},
    nodes near coords,
  ]
  \addplot[style={blue,fill=blue,mark=none}] coordinates { (3849,Original)         (47,LUT)
 };
  \addplot[style={red,fill=red,mark=none}] coordinates { (3020,Original)         (4,LUT)
 };
    \addplot [style={green!40!gray,fill=green!40!gray,mark=none}]coordinates { (26,Original)         (0,LUT)
  };
  \legend{LUT, FF, DSP slices}
  \end{axis}
\end{tikzpicture}
\caption{LUT approximation}
\label{fig:complexity_LUT}
\end{subfigure}
\caption{Tanh implementation complexity in terms of LUT, FF, and DSP slices for the Taylor series, PWL, and LUT approximations after the Xilinx realization pipeline.}
\label{fig:resources_approx}
\end{figure*}

\subsection{Quality of Transmission: Improvement Study}
Fig.~\ref{fig:performance_result} summarizes the performance of the NN-based equalizers compared to 1 StpS DBP and CDC over different launch powers for simulated and experimental data. 
The results referring to the simulated data are given in Fig.~\ref{fig:performance_result_sim}. The biLSTM equalizer shows approximately the same performance as a 1 StpS DBP while improving the optimal power from $-$1 dBm to 2~dBm and the Q-factor by 1.3 dB with respect to the CDC. Regarding deep CNN, it performs worse than the biLSTM and the 1-StpS DBP; the optimal power is improved from $-$1 dBm to 1~dBm and the Q-factor increases by 0.8~dB compared to the CDC. On the other hand, with the experimental data\footnote{The Q-factor obtained with the Python model and the FPGA implementation were virtually identical because we did not consider quantization.}, we observe in Fig.~\ref{fig:performance_result_exp} that the biLSTM outperforms the 1-StpS DBP, especially in the noise-dominated region. For the 1-StpS DBP case with the experimental data, the Q-factor increases by 1.3~dB in the simulation and by 1.5~dB in the experiment. Compared to simulation, NN-based equalizers in the experiment also lead to a higher gain for the Q-factor; when having CDC as a baseline, the gain improves from 1.3~dB (simulation) to 1.7~dB (experiment) in the case of biLSTM equalizers, and from 0.8~dB (simulation) to 1~dB (experiment) for the deep CNN. The optimal power also shifts from 0~dBm to 1~dBm in both cases. This shows that the NN has the potential to reduce the effects of both the Kerr nonlinearity and the component-induced corruptions which can be the effects of the transceivers (ADC/DAC or drive amplifier) and other effects that are not considered in the simulation such as some polarization mismatch, connector loss, different fiber parameters along the fiber, for both Tx and Rx sides. 

In fact, all component impairments in the simulations were modeled with white noise (to represent the non-considered impairments from a real transmission and the non-ideal transceivers), so the equalizer could not mitigate them deterministically, whereas in the experimental case, our equalizer could enhance the Q-factor slightly further. Numerically, this can be observed by the fact that there is a 3dB enhancement of the optimal launch power compared to the CDC in the simulation and only 1dB in the experiment. This can be explained because, in the experiment, other effects apart from the Kerr effects were also compensated more in the linear regime, so a higher Q-factor was achieved but with a lower launch power. However, the maximum Q-factor after equalization in the simulation (5.24dB at 2dBm) is lower than the one achieved in the experiment for the same launch power (5.54dB at 2dBm). Hence, more linear impairments than nonlinear ones are recovered in the experiment, resulting in a smaller improvement in launch power.
In particular, the biLSTM equalizer beats the deep CNN equalizer because the biLSTM is a recurrent-based NN that benefits from temporal sequential data learning\cite{freire2021performance,deligiannidis2020compensation}.

\subsection{Nonlinear Activation Function: Performance versus Complexity Investigation} \label{sec:result_act}
Now, having obtained the Q-factor benchmarks for the NN-based equalizers, we move on to the investigation of performance, studying different approximation techniques for nonlinear activation functions: Taylor series, PWL, and LUT. Fig.~\ref{fig:performance_vs_comp} depicts the Q-factor in the optimal power after equalization for three scenarios: the original NN without approximation, the NN with approximation (without re-training), and the NN with approximation (with re-training). Note that training the NN from scratch when replacing exact activation functions with approximations takes a considerably longer time to converge than retraining the original NN after replacing floating-point activation functions with their approximations. The training from scratch with the Taylor and LUT activations approximation even results in lower eventual performance. Therefore, in this figure, we only report the results of the retraining approach. Fig.~\ref{fig:performance_vs_comp_taylor} and Fig.~\ref{fig:performance_vs_comp_PWL}, corresponding to the Taylor series and PWL, respectively, reveal the same trend. Without training, as the complexity of the approximation increases, the NN equalizer performs clearly better. However, with training, our increasing complexity barely improves the performance: the NN is able to adjust its parameters to mitigate the approximation error and provides comparable performance to the NN without approximations. The Q-factor versus complexity (order) of approximation plots,  Fig.~\ref{fig:performance_vs_comp}, highlight the remarkable performance gain in all considered approaches when we re-train the model with activation functions replaced by the approximation, and we see that training can mitigate the errors from the approximation. This means that even the low-order approximations, such as the simplest PWL with three segments, can still yield results nearly identical to those rendered by the original activation functions.

Fig.~\ref{fig:performance_vs_comp_LUT} shows the performance of the LUT approximation. When replacing the activation functions with LUT without re-training, a certain number of bits is needed to provide an acceptable Q-factor level
\footnote{Note that in this study, we followed the LUT implementation from Ref. \cite{yang2018design,namin2009artificial} which presented the LUT with equal x-error intervals. 
The alternative approach (activation functions with equal y-error intervals) can be used, but in our case, there is only a slight improvement in the Q-factor when the number of bits is greater than 5 and with the re-training, the performance is very close to the x-interval approach.}
For example, the minimum number of bits needed to provide a Q-factor greater than zero is 7 bits; 9 bits are needed to provide performance comparable to the model without approximation. On the other hand, when re-training the NN after the approximation, the Q-factor for the lower number of bits (from 3 to 7 bits) considerably increases. In this case, the non-differentiability makes the training challenging and limits the performance reachable in training, but the improvement is still noticeable when the number of bits is between 3 and 7. Fig.~\ref{convergency_AF} shows the convergence speed of the three approximation techniques. It can be seen that the learning of Taylor and PWL is similar, whereas the re-training of LUT approximation is more difficult. Although the LUT gradient, Fig.~\ref{fig:derivative_LUT}, and the PWL gradient in Fig.~\ref{fig:derivative_PWL} seem interchangeable, the forward propagation of the LUT approximation is still discrete, which means that with the lower number of bits we create a large gap between each quantized level. Thus, small changes that the gradient makes to update the weights might not change the quantization level to the next value. This means that the loss region is the same as it was in the last NN training interaction (trapped in a local minimum). Notably, in~\cite{yin2019understanding} a similar circumstance was observed; the previous reference also pointed to the instability of the training that can occur with a quantized activation function. In the case of PWL, the learning is more stable due to the continuity of the function's approximation, as each weight update generates a new loss value and a distinct point in the forward propagation.

In addition, as anticipated, we observe that when quantizing the LUT below 4 bits, no acceptable Q-factor can be reached even after the re-training. The reason for this is that when we quantize the activation function, unlike when we quantize the weights, we are limited in our ability to represent the modulation of the equalized signal. In our situation, we use 16QAM, which requires at least 4 bits to represent a constellation data point. However, as we see, even 4 bits are insufficient in this case to preserve all the essential features for the equalization process when using the quantization of the activation function. 

When more bits are used, a better Q-factor can be achieved; however, more memory is then required to represent the quantization. It is worth noticing that when the number of bits is greater than 10, the Q-factor no longer improves in both scenarios (with and without re-training).

The amount of resources required (in terms of LUT, FF, and DSP slices) in the FPGA when using the approximations for tanh, is compared to that when applying the actual tanh activation function in Fig.~\ref{fig:resources_approx}. This figure depicts the resources used to build the logic behind the functionality of each approximation. Note that the coefficients and values used in each panel of Fig. \ref{fig:acti_func_fpga}, are considered an input of the implemented box, which is accessed by the FPGA memory. The implementation complexity of the actual float activation functions is significantly higher than that of the approximations. In the Taylor series approximations, Fig.~\ref{fig:complexity_taylor}, the number of FF and LUT used to implement the approximations is drastically reduced compared to the original activation functions; to be more specific, when the highest order of the polynomial is 9, the number of FF required decreases by 6.7 times, and the number of LUT required is three times smaller. In terms of DSP slices, the 9$^{th}$ order approximation requires 6 DSP slices fewer than the original functions. As the approximation becomes simpler, the implementation requires fewer resources, as expected. For the PWL approximation, no usage of DSP slices is required for the implementation. Like in the Taylor series approximation, the number of FF and LUT required decreases noticeably. Compared to the original float-precision activation function, the PWL with 9 segments requires 2.8 times less LUT, and 13 times less FF. As the complexity decreases according to the number of segments, fewer resources are needed. Turning to the LUT approximation, it does not require any of the DSP slices as well, and the number of LUT and FF decreases by a factor of 80 and 775, respectively. Regardless of the number of bits in the quantized activation function, approximately the same amount of resources is required to implement the logic of the LUT approximation (see Fig.\ref{fig:complexity_LUT}). The LUT approximation approach is an algorithm based on evaluating the closest value in the LUT from a certain input and determining the memory address index that corresponds to that closest value to retrieve the information. As the number of bits increases, a larger memory is needed to store the LUT approximation points, which are a quantized version of the function. However, we do not account for this memory usage in our study because this is considered one of the inputs to our implemented box.

In conclusion, when performance, memory, and resources are considered, the PWL emerges as a viable candidate for hardware implementation, particularly, the 3-segment PWL variant with re-training. When the model learns to reduce approximation errors, the Q-factor of 3-segment PWL can reach a level comparable to that of the original activation functions; in addition, there is no need for DSP slices, resulting in more efficient use of resources than the Taylor approximation. With this, the RAM usage in the PWL is efficient because only a few coefficients of the approximation must be saved, whereas the LUT, which brings about difficulties during the re-training process, requires that all values of each quantization level be saved, resulting in an exponential increase in memory usage as the number of bits increases.

\subsection{Computational Complexity Analysis for NN Equalizers versus CDC, Implemented in Different Platforms}\label{sec:CCvsCDC}
\subsubsection{Standard FPGA implementation (with DSP slices)}

\begin{figure*}[th]
\begin{subfigure}{.33\textwidth}
\centering
  \includegraphics[scale=0.7]{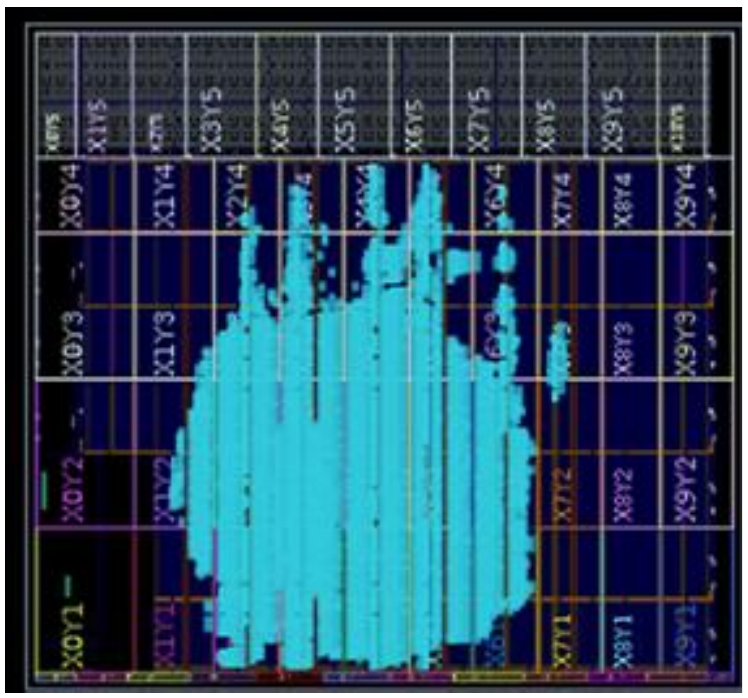}
\caption{biLSTM eq.}
\label{fig:results_chip_bilstm}
\end{subfigure}\hfill
\begin{subfigure}{.33\textwidth}
\centering
  \includegraphics[scale=0.7]{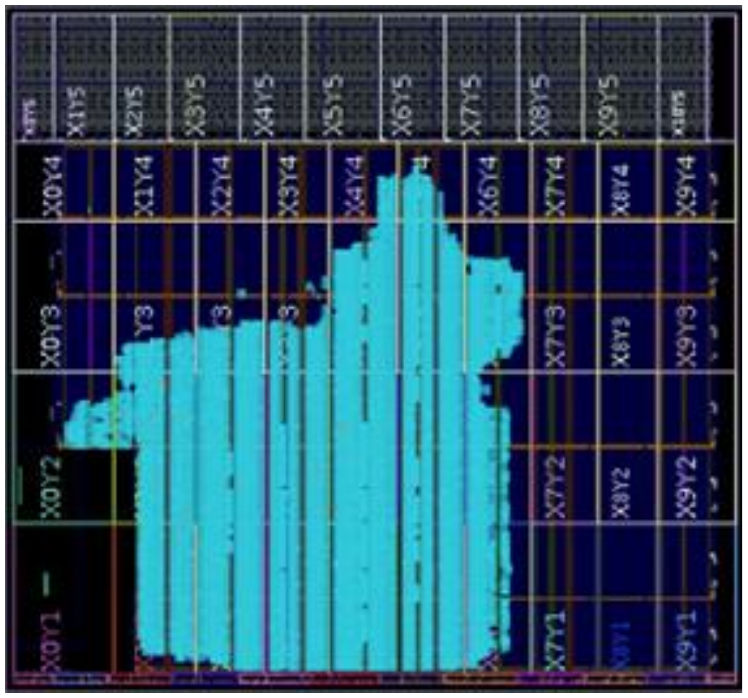}
\caption{Deep CNN eq.}
\label{fig:results_chip_cnn}
\end{subfigure}\hfill
\begin{subfigure}{.3\textwidth}
\centering
  \includegraphics[scale=0.7]{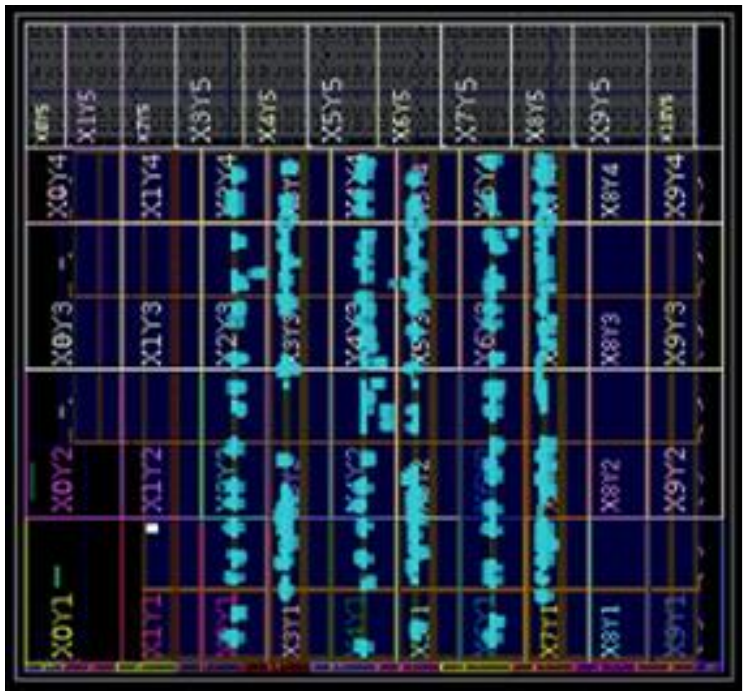}
\caption{CDC block}
\label{fig:results_chip_cdc}
\end{subfigure}
    \caption{Implementation in the EK-VCK190-G-ED Xilinx FPGA~\cite{FPGA} of the (a) biLSTM eq., (b) Deep CNN eq., (c) CDC block - Time Domain.}
    \label{fig:results_chip}
\end{figure*}

Figs.~\ref{fig:results_chip_bilstm}, \ref{fig:results_chip_cnn}, and \ref{fig:results_chip_cdc} show the real implementation and chip areas used for biLSTM, deep CNN equalizers, and CDC, respectively, on the state-of-the-art EK-VCK190-G-ED Xilinx FPGA chip\cite{FPGA}  VCK190 kit features an AMD Xilinx Versal ACAP VC1902-2. \footnote{Note that ``C'' identifies it as a core series device and  ``-2'' indicates the middle-speed grade.}. The device has 1968 DSP engines, 1799680 FFs, 899840 LUTs, and 400 AI engines. In this paper, the AI engines are not used, as the HLS tool used in this work does not support the AI cores. The AI engines in the Versal AI Core FPGA are specialized units optimized for machine learning workloads and are not directly exposed to the programmer via the Vitis HLS tool\footnote{Pre-built AI libraries and IPs such as the Xilinx Deep Learning Processor (xDNN) library, which is optimized for DNNs, can be used 
 for running on the FPGA's AI Engines. Note that the Vitis HLS tool can be used in conjunction with the xDNN library to target AI cores resources on Xilinx FPGAs\cite{Xilinx_vitis_tutorial}, but this is outside our current scope.}. This chip is partitioned into 40 clock regions, with the blue areas in Fig.~\ref{fig:results_chip} representing the used chip resources. Table~\ref{tab:vck_190_implement} summarizes the most important information on the VCK 190 implementation of the biLSTM, deep CNN equalizers, and the CDC, in terms of latency, clock frequency\cite{FPGA}, resources required, the utilization of the resources, and throughput. The achieved throughput ($\mathrm{T_{a}}$) can be calculated as follows:
\begin{equation}
    \mathrm{T_{a}} =  \mathrm{clock}\times\log_2(\mathrm{QAM})\times n_\text{out},
\end{equation}
where clock is the clock frequency, QAM is the modulation format, $\log_2(\mathrm{QAM})$ is the number of bits per symbol, and $n_\text{out}$, is the number of parallel symbols we recover in the output; in our case, $n_\text{out} = 61$.

Three important conclusions can be drawn from that figure. First, although the biLSTM renders a higher Q-factor improvement, due to the equalizer's recurrent structure its feedback loop connections cause a bottleneck in the design, resulting in higher latency (33 $\mu$s) and lower clock frequency (270 MHz). On the other hand, deep CNN and CDC can be parallelized more efficiently. The parallelizability brings about a reduction in their latency to 19.9 $\mu$s for the deep CNN, and 1.1 $\mu$s for the CDC. Due to the fact that the CDC has one filter, whereas the deep CNN has 70 filters, the parallelization is easier in the CDC implementation because of hardware restrictions, resulting in an operating frequency of 524 MHz for the CDC case, and 244 MHz for the deep CNN case. Fig.~\ref{fig:results_chip_cdc} clearly shows the CDC parallelization.  A long latency increases the time required to process each time step, leading to slower overall processing times and reducing the speed of the network. This can be problematic in real-time applications where a fast response is necessary. To mitigate the impact of long latency, design optimization techniques can be applied to reduce the latency and increase the performance, such as reducing the size of the memory blocks, using more efficient algorithms, and implementing pipelining. In this work, due to our offline processing consideration, all the input is already in the memory, and the request of the sequence with 81 symbols as inputs is created so that the functioning of the NN-based equalizer parallelization works. 

Second, regarding the FPGA utilization, the biLSTM equalizer is the only one using Block Random Access Memory (BRAM)\footnote{BRAM is a type of memory in FPGAs that is used to store large amounts of data, typically used in FPGA designs to implement memory-intensive functions such as image and video processing, buffers, and large arrays. Unlike other memory elements in an FPGA, BRAM is a dedicated memory that is separate from the FPGA's general-purpose FFs and LUTs.}
 to store future/past recurrent states, while both CNN and CDC do not need such blocks. BRAM is used to store the hidden states of an LSTM, which can then be fed back into the network at the next time step to maintain its memory, while in the case of the feedforward structures, the special LUTs are used to store the coefficients. The structure of the system of an LSTM cell is shown in Fig. \ref{fig:lstmcell}, and the buffer used in the implementation was synthesized as BRAMs as global memory on the chip \cite{he2021fpga}.
  Note that the number shown in the tables is the number of BRAM blocks, which was the automatic result reported after the Synthesis step (Vivado). By using BRAM, the hidden state information can be stored in a dedicated memory block, separate from the other resources in the FPGA. This can lead to improved performance, as memory accesses are optimized and dedicated resources are used for memory storage. However, the size of the BRAM blocks and the memory requirements for the recurrent connections should be carefully considered when designing an LSTM on an FPGA. The available BRAM resources may be limited, and it may be necessary to trade off memory size for performance, depending on the requirements of the specific LSTM design. 
 The SRL (Shift Register Look Up Table) in Table \ref{tab:vck_190_implement} and Table \ref{tab:imple_only_LUT_FF} is a mode available in FPGAs whereby the LUT-RAM is configured as a shift register structure. This is more efficient, as it requires fewer cells and less routing than using individual DFFs to build shift registers.
For the usage of DSP slices, LUT, and FF in each equalizer type, the biLSTM requires 64$\%$ DSP slices and 13$\%$ of LUT and FF, the deep CNN uses 30$\%$ DSP slices, 13$\%$ of LUT and 21$\%$ of FF, and the CDC needs 54$\%$ DSP slices and 1$\%$ of LUT and FF. 

Third, in terms of throughput, the clock frequency is the maximum that each implementation can handle to comply with a zero-negative slack design. In this sense, the total throughput for the 16QAM modulation format is 66G, 60G, and 127G, for the biLSTM, deep CNN, and CDC block, respectively. 

\begin{table*}[tbh]
   \centering
   \resizebox{1.021\textwidth}{!}{
\begin{tabular}{c c c c c c c c c c c c c }
\hline
Type    & \begin{tabular}[c]{@{}c@{}}Latency\\ ($\mu s$)\end{tabular} & \begin{tabular}[c]{@{}c@{}}Clock\\ Frequency\\ (MHz)\end{tabular} & BRAM & SRL & \begin{tabular}[c]{@{}c@{}}DSP \\ Slices\end{tabular} & LUT & FF  & \begin{tabular}[c]{@{}c@{}}Utilization (\%)\\ $[$ DSP/LUT/FF $]$\end{tabular}& \begin{tabular}[c]{@{}c@{}}Throughput\\ ($Gbits/s$)\end{tabular}  & \begin{tabular}[c]{@{}c@{}}$N^{o}$ FPGAs\\ for 200G\end{tabular}& \begin{tabular}[c]{@{}c@{}}$N^{o}$ FPGAs\\ for 400G \\ $[$dual-carrier$]$\end{tabular}& \begin{tabular}[c]{@{}c@{}}$N^{o}$ FPGAs\\ for 400G \\ $[$56GBd$]$\end{tabular} \\ \hline\hline
\myrowcolour
biLSTM+CNN & 33.4                                                    & 270                                                               & 164  & 109 & 1260                                                   & 113532 & 224386 & \textcolor{red}{64.0}/12.6/12.5 & 66                                                            & 2.6 & 5.3  & 7.2                                                        \\ 

Deep CNN & 19.9                                                    & 244                                                               & 0  & 125 & 582    & 118477                                                & 379829 & \textcolor{red}{29.6}/13.2/21.1 & 60                                                            & 1.4& 2.7& 3.7                                                    
\\
\myrowcolour%
CDC     & 1.1                                                    & 524                                                               & 0  & 1 & 1072 & 10441 & 5640 & \textcolor{red}{54.5}/1.2/0.3 & 127                                                            & 1.2 & 2.3 & 3.1                                                        \\ \hline

\end{tabular}}
    \caption{VCK190~\cite{FPGA} Implementation.}
    \label{tab:vck_190_implement}
\end{table*}

\begin{table*}[ht!]
    \centering
\begin{tabular}{c c c c c c c c c c c c}
\hline
Type    & \begin{tabular}[c]{@{}c@{}}Latency\\ ($\mu s$)\end{tabular} & \begin{tabular}[c]{@{}c@{}}Clock\\ Frequency\\ (MHz)\end{tabular} & BRAM & SRL & LUT & FF  & \begin{tabular}[c]{@{}c@{}}Utilization (\%)\\ $[$ LUT/FF $]$\end{tabular}& \begin{tabular}[c]{@{}c@{}}Throughput\\ ($Gbits/s$)\end{tabular}  & \begin{tabular}[c]{@{}c@{}}$N^{o}$ FPGAs\\ for 200G\end{tabular} & \begin{tabular}[c]{@{}c@{}}$N^{o}$ FPGAs\\ for 400G \\ $[$dual-carrier$]$\end{tabular}& \begin{tabular}[c]{@{}c@{}}$N^{o}$ FPGAs\\ for 400G \\ $[$56GBd$]$\end{tabular} \\ \hline\hline
\myrowcolour
biLSTM+CNN & 39.2                                                    & 234                                                               & 164  & 163                                                  & 566070 & 249763 & \textcolor{red}{62.9}/13.9 & 57                                                             &         3.0  &          6.0  &          8.1                                              \\ 

Deep CNN & 17                                                   & 245                                                               & 0  & 162   & 300426                                                & 399868 & \textcolor{red}{33.4}/22.3 & 60                                                             &   1.5     &         3.0  &          4.1                                             
\\
\myrowcolour%
CDC     & 2.3                                                    & 246                                                              & 0  & 1 & 418534 & 24968 & \textcolor{red}{46.5}/1.4 & 60                                                           & 2.1 &          4.2  &          5.7                                                  \\ \hline

\end{tabular}
    \caption{Implementation where all multiplications are done using LUT and FF.}
    \label{tab:imple_only_LUT_FF}
\end{table*}

Lastly, regarding the calculation of the number of equivalent FPGAs for a certain target throughput ($\mathrm{T_{target}}$) from an experiment, we have considered the following equation: 

\begin{equation}
    \mathrm{N_{FPGA}}= \frac{\mathrm{T_{target}}}{\mathrm{T_{a}}} * U_t,
\end{equation}

 where $\mathrm{T_{a}}$ is the throughput achieved after the NN design pipeline, and $U_t$ is the maximum utilization after the NN design pipeline (both reported in Table~\ref{tab:vck_190_implement}). In the CNN+biLSTM case, for example, because the experiment was 16QAM single carrier transmission at both 34 GB per pol, the target throughput is 272 Gbit/200 G. So, considering the maximum utilization of 64\% and the throughput achieved of 65.9Gbit by the NN equalizer, $\mathrm{N_{FPGA}}$ would be equal to $2.6$ FPGAs.

 In fact, we considered the FPGA estimation for three different cases:

 \begin{enumerate}
    \item 200G scenario: which is the scenario of the experiment in this paper -  16QAM single carrier configurations at both 34GBd per pol (resulting in 272Gbit/200G).
    \item 400G scenario: considering a dual carrier transmission instead of a single carrier. In this case, we just need to scale the resources of 200G by a factor of 2.
    \item 400G scenario: considering a 16QAM single carrier configurations with higher symbol rate equal to  56GBd per pol (resulting in 448Gbit, with 12\% FEC overhead). In this case, we simply multiply the 200G resources by $56^2/34^2 = 2.71$ because, given the increased symbol rate, the resources will grow approximately quadratically with the increase in symbol rate since our implementations were in the time domain. 
\end{enumerate}

For case 1, we observe that 200G transmission can be achieved using an equivalent FPGA that has the same capacity as $\approx$ 3~FPGAs (VCK190) in the case of biLSTM, $\approx$ 2~FPGAs in a deep CNN case, and $\approx$ 1~FPGAs for CDC.
For case 2, 400G with dual carrier transmission can be achieved using an equivalent FPGA that has the same capacity as $\approx$ 5~FPGAs (VCK190) in the case of biLSTM, $\approx$ 3~FPGAs in a deep CNN case, and $\approx$ 2~FPGAs for CDC. Finally, in case 3, because of the increase in symbol rate, much more hardware was needed. For the 400G with 56Gbd transmission case, we would need an equivalent FPGA that has the same capacity as $\approx$ 7~FPGAs (VCK190) in the case of biLSTM, $\approx$ 4~FPGAs in a deep CNN case, and $\approx$ 3~FPGAs for CDC. In all three cases, biLSTM used approximately 2.5 times more FPGA than required by the CDC implementation.

\subsubsection{ASIC Equivalent Implementation (no DSP slices)}

Unlike FPGAs, ASICs are application specific, and their digital circuitry contains permanently connected gates and FF in silicon; therefore, in ASIC design, there is no configurable block (such as DSP blocks). In this subsection, we evaluate the approximations of the resource requirements and the performance in terms of throughput, clock frequency, and latency of the ASIC implementation by considering the VCK190 implementation without the usage of DSP slices. Table~\ref{tab:imple_only_LUT_FF} contains information for the implementation of the biLSTM, the deep CNN, and the CDC equalizer on the FPGA with only LUT and FF. The biLSTM and CDC have a noticeably higher latency: 5.8 $\mu$s and 1.2 $\mu$s higher, respectively, compared to implementation with DSP slices detailed in Table~\ref{tab:vck_190_implement}. The lower clock frequency is also observed: 234 MHz for biLSTM and 246 MHz for the CDC, resulting in a lower throughput: 57G for biLSTM and 60G for the CDC. The degradation in throughput, latency, and clock frequency highlights the fact that DSP slices speed up the execution of signal processing functions. Especially in the CDC, when we allow implementation with DSP slices, the number of LUT and FF used is 40 times and 4.4 times less, respectively. Therefore, we can clearly see the degradation of performance in terms of the throughput of the CDC when the DSP slices are not used. However, in the case of deep CNN, the latency decreases by 2 $\mu$s, the clock frequency increases by 1 MHz, and the throughput remains unchanged. We can observe that in the deep CNN implementation with the DSP slices in Table~\ref{tab:vck_190_implement}, the number of DSP slices used is only about half that for the other two equalizers, and the number of LUT and FF is highest. Therefore, our removal of DSP slices did not affect the deep CNN because the processing time for LUT and FF was already the bottleneck in the previous implementation.
Here, it is essential to recall that the clock frequency, which is essential to establishing the throughput, is chosen to guarantee zero negative timing slack. In this regard, since the routing and mapping are performed automatically by the Vivado platform, we can observe that when restricting the software from using the DSP slices, longer paths are generated during the synthesis in the biLSTM and CDC cases. The longer paths cause the clock frequency to decrease to achieve the zero-negative slack level. In contrast, the deep CNN's paths when deploying the DSP slices are already long due to the logic implementation and synthesis, and, therefore, there is no significant variation in clock frequency after removing the DSP slice in that case.
 
Regarding the utilization of LUT and FF: the biLSTM uses 62.9\% of LUT and 13.9\% of FF, the deep CNN uses 33.4\% of LUT and 22.3\% of FF, and the CDC uses 46.5\% of LUT and 1.4\% of FF. The utilization of LUT for all three equalizers is considerably increased compared to the previous case (standard FPGA implementation). As the number of LUTs and FFs increases, the equivalent number of FPGAs used to represent the biLSTM and the CDC equalization also grows. 

For the same three cases we have discussed previously, we could observe the following. For case 1, 200G transmission can be achieved using an equivalent FPGA that has the same capacity as $\approx$ 3~FPGAs (VCK190) in the case of biLSTM, $\approx$ 2~FPGAs in a deep CNN case, and $\approx$ 2~FPGAs for CDC.
For case 2, 400G with dual carrier transmission can be achieved using an equivalent FPGA that has the same capacity as $\approx$ 6~FPGAs (VCK190) in the case of biLSTM, $\approx$ 3~FPGAs in a deep CNN case, and $\approx$ 4~FPGAs for CDC. Eventually, in case 3 with a 16 QAM 56Gbd transmission, we would need an equivalent FPGA that has the same capacity as $\approx$ 8~FPGAs (VCK190) in the case of biLSTM, $\approx$ 4~FPGAs in a deep CNN case, and $\approx$ 6~FPGAs for CDC. In all three cases, biLSTM used approximately 1.5 times more FPGA than required by the CDC implementation.

Finally, we note that in this study, we established the approximate resources and performance of the equalizers implemented on ASIC by excluding the DSP slices. However, this is still not an optimized realization: in ASIC implementation, the number of resources used needs to be further optimized to reduce the utilization, increase the clock frequency, and enable high-speed processing.

\section{Conclusion and Open Challenges} \label{sec:open_challenges}

In this paper, we carry out a detailed study of the design of NN-based optical equalizers, addressing the steps from Python realization to FPGA implementation. To approach the real hardware implementation of NN-based equalizers, we investigated three approximation approaches (Taylor, piecewise linear, and lookup table) for nonlinear activation functions, aiming at reducing the computational complexity. The complexity, performance, and resources required for the approximations have been evaluated. We then examined the biLSTM equalizer implementation on the FPGA, assessing the complexity reduction due to the implementation using fixed-point arithmetic and nonlinear activation function approximations. Our realization showed that the biLSTM requires only $\approx$ 2.5 times more FPGA resources than the CDC implemented in the time domain, while still outperforming the 1-StpS DBP in Q-factor.
The approximate utilization of ASIC when using only FF and LUT to implement the logic of 
such DSP blocks for channel equalization were also considered. The results obtained for the ASIC estimation showed a drop in throughput for the biLSTM equalizer, due to the challenges in route design to achieve zero negative slack using higher than 234~Mhz clock frequencies. The latter indicates that, for future applications, much more effort is still required in this direction.

We consider this work to be yet another piece of evidence that the deployment of NN-based equalizers in commercial applications might become a reality. Indeed, it is already quite clear that the NN-based equalizers can provide significant performance improvements when implemented on top of the existing DSP algorithms. The NN can even replace some DSP chain parts, such as the CDC block. Moreover, as we evaluated in this work, the real-time hardware implementation of NNs is already an attainable reality. Unfortunately, the complexity of the proposed implementation is still too high for NN equalization deployment in commercial optical coherent transponders. 
Additionally, one of the most critical aspects of the DSP block of a coherent transponder is its power efficiency. Since power efficiency is a direct consequence of the complexity, further investigations focusing on the reduction of NN's complexity (when those are specifically implemented in hardware) are important. These investigations should address several areas, including the simplification methods for NN structures, such as pruning, weight sharing, quantization, and the respective hardware implementation aspects.

To reduce complexity, approaches concentrating on different transmission scenarios and, consequently, adopting different DSP systems, may be envisioned. Core and regional optical networks are characterized by quite long optical links, where optical signals experience noticeable nonlinear distortions and the accumulated chromatic dispersion is large enough. Access and metro optical networks are characterized by short propagation distances, in which the predominant transmission effects are typically the ASE noise and limited optical power at the RX input (e.g., in the point-to-multipoint solutions). An important standpoint from the industry is how the implemented NN (particularly, the simplification and complexity reduction strategies) varies with the change of transmission scenario. If the low-complexity implementation of a given NN works well in one situation but not in another, this can pose a serious difficulty, as we often cannot afford to produce a unique chip for every circumstance. It is desirable that the NN after pruning and quantization is still capable to equalize versatile transmission setups (working, e.g., for different fiber types, span lengths, launch powers, etc.)

Finally, we list the unresolved questions that were not investigated in our work but can be crucial for further research. 

\begin{itemize}
    \item Power consumption evaluation of reduced-complexity NN equalizers.
    \item How to realize a NN that can work for multiple transmission scenarios with no or very limited retraining.
     \item Parallelization of the recurrent NN structures study.
     \item Implementation in the FPGA of more robust quantization levels moving from int32, as presented in this manuscript, to int8 or less, if possible, by using heterogeneous quantization together with quantization-aware training\cite{freire2022reducing}.
     \item Further flexibility analysis to avoid the need to retrain the hardware NN implementation, e.g., the hardware tests of domain adaptation/randomization and transfer learning \cite{freire2021transfer,freire2022domain}.
\end{itemize}

\begin{appendices}

\begin{table*}[ht]
\centering
\begin{tabular}{cccccccc}
\hline
                                  & \multicolumn{7}{c}{Functions}                                                                                                                                                                                                                                                                                                                                                                                                                                                                                                                                                                                                                                                                                                                                            \\ \cline{2-8} 
                                  & \multicolumn{3}{c}{tanh}                                                                                                                                                                                                                                                                                                                                                      &  & \multicolumn{3}{c}{sigmoid}                                                                                                                                                                                                                                                                                                                                                           \\ \cline{2-8} 
\multirow{-3}{*}{No. of segments} & Equation                                                                                                                                                                           &  & Condition                                                                                                                                                                             &  & Equation                                                                                                                                                                                           &  & Condition                                                                                                                                                                     \\ \hline
\rowcolor[HTML]{EFEFEF} 
3                                 & \begin{tabular}[c]{@{}c@{}}$1$\\ $0.90909x$\\ $-1$\end{tabular}                                                                                                                    &  & \begin{tabular}[c]{@{}c@{}}$x>1.1$\\ $-1.1<x\le 1.1$\\ $x\le-1.1$\end{tabular}                                                                                                        &  & \begin{tabular}[c]{@{}c@{}}$1$\\ $0.22727x+0.5$\\ $0$\end{tabular}                                                                                                                                 &  & \begin{tabular}[c]{@{}c@{}}$x>2.2$\\ $-2.2<x\le 2.2$\\ $x\le-2.2$\end{tabular}                                                                                                \\
5                                 & \begin{tabular}[c]{@{}c@{}}$1$\\ $0.41666x+0.29166$\\ $x$\\ $0.41666x-0.29166$\\ $-1$\end{tabular}                                                                                 &  & \begin{tabular}[c]{@{}c@{}}$x>1.7$\\ $0.5<x\le 1.7$\\ $-0.5<x\le0.5$\\ $-1.7<x\le-0.5$\\ $x\le-1.7$\end{tabular}                                                                      &  & \begin{tabular}[c]{@{}c@{}}$1$\\ $0.17223x+0.55219$\\ $0.23747x+0.5$\\ $0.17223x+0.44781$\\ $0$\end{tabular}                                                                                       &  & \begin{tabular}[c]{@{}c@{}}$x>2.6$\\ $0.8<x\le 2.6$\\ $-0.8<x\le0.8$\\ $-2.6<x\le-0.8$\\ $x\le-2.6$\end{tabular}                                                              \\
\rowcolor[HTML]{EFEFEF} 
7                                 & \begin{tabular}[c]{@{}c@{}}$1$\\ $0.285x+0.48699$\\ $0.57214x+0.17114$\\ $x$\\ $0.57214x-0.17114$\\ $0.285x-0.48699$\\ $-1$\end{tabular}                                           &  & \begin{tabular}[c]{@{}c@{}}$x>1.8$\\ $1.1<x\le 1.8$\\ $0.4<x\le1.1$\\ $-0.4<x\le0.4$\\ $-1.1<x\le-0.4$\\ $-1.8<x\le-1.1$\\ $x\le-1.8$\end{tabular}                                    &  & \begin{tabular}[c]{@{}c@{}}$1$\\ $0.12363x+0.62909$\\ $0.18701x+0.54036$\\ $0.23747x+0.5$\\ $0.18701x+0.45964$\\ $0.12363x+0.37091$\\ $0$\end{tabular}                                             &  & \begin{tabular}[c]{@{}c@{}}$x>3$\\ $1.4<x\le 3$\\ $0.8<x\le1.4$\\ $-0.8<x\le0.8$\\ $-1.4<x\le-0.8$\\ $-3<x\le-1.4$\\ $x\le-3$\end{tabular}                                    \\
9                                 & \begin{tabular}[c]{@{}c@{}}$1$\\ $0.14331x+0.68417$\\ $0.3381x+0.412$\\ $0.269382x+0.09185$\\ $x$\\ $0.269382x-0.09185$\\ $0.3381x-0.412$\\ $0.14331x-0.68417$\\ $-1$\end{tabular} &  & \begin{tabular}[c]{@{}c@{}}$x>2.2$\\ $1.4<x\le 2.2$\\ $0.9<x\le 1.4$\\ $0.3<x\le0.9$\\ $-0.3<x\le0.3$\\ $-0.9<x\le-0.3$\\ $-1.4<x\le-0.9$\\ $-2.2<x\le-1.4$\\ $x\le-2.2$\end{tabular} &  & \begin{tabular}[c]{@{}c@{}}$1$\\ $0.08514x+0.71051$\\ $0.12644x+0.62791$\\ $0.182242x+0.09185$\\ $0.23747x+0.5$\\ $0.08514x+0.45585$\\ $0.12644x+0.37209$\\ $0.182242x+0.28949$\\ $0$\end{tabular} &  & \begin{tabular}[c]{@{}c@{}}$x>3.4$\\ $2<x\le 3.4$\\ $1.5<x\le 2$\\ $0.8<x\le1.5$\\ $-0.8<x\le0.8$\\ $-1.5<x\le-0.8$\\ $-2<x\le-1.5$\\ $-3.4<x\le-2$\\ $x\le-3.4$\end{tabular} \\ \hline
\end{tabular}
\caption{PWL approximation equations of sigmoid and tanh for 3, 5, 7 and 9 segments.}
\label{tab:piecewise}
\end{table*}

\section{FPGA Notation}
\label{app:fpga}
Offline FPGA implementation refers to the process of designing and configuring an FPGA before it is deployed in a target system. This typically involves using specialized software tools to design, simulate, and verify the digital logic that will be implemented on the FPGA. The final design is then converted into a format that can be loaded onto the FPGA, such as a bitstream file.

A D-type Flip-Flop (DFF) is a type of sequential logic element that is commonly deployed in digital systems as they are simple. DFFs can be found in digital state machines, shift registers, counters, and other digital circuits that require memory storage.  They are often used to store the value of a digital signal and can be used  with other logic elements, such as AND and OR gates, to build more complex digital systems.

BRAM (Block Random Access Memory) is a type of memory available in FPGAs that is used to store data. Unlike other memory elements in an FPGA, BRAM is dedicated memory that is separate from the FPGA's general-purpose flip-flops and LUTs. BRAM is used to store large amounts of data, typically used in FPGA designs to implement memory-intensive functions such as image and video processing, buffers, and large arrays.

RAM-based LUT (LUT RAM) or distributed RAM is sometimes called in the user guides as the RAM used to store the logic function equations. LUT RAM can also be configured to be used as user storage with a similar function to BRAM.

Digital Signal Processing (DSP) Blocks or Slices: Digital Signal Processing (DSP) blocks or slices are specialized components within an FPGA that are designed specifically for processing digital signals. They contain dedicated hardware resources such as multipliers, adders, accumulators, and registers that can perform complex mathematical operations at high speeds. They are optimized for efficient use of resources and can perform operations in parallel, which enables high-speed processing of large volumes of data. Additionally, DSP blocks are typically designed to support fixed-point and floating-point arithmetic, and they can be configured to support various data widths and precision. They can also be combined with other components within an FPGA to create complex signal processing pipelines.

\section{PWL equations}\label{app:act_func}
The equations of the PWL approximations of sigmoid and tanh can be found in Table~\ref{tab:piecewise} for 3, 5, 7, and 9 segments.

\end{appendices}

\Urlmuskip=0mu plus 1mu\relax
\bibliographystyle{IEEEtran}
\bibliography{references}

\end{document}

%% file: convergence_speed.tikz
\begin{tikzpicture}[scale=0.6]
    \begin{axis} [, 
        xlabel={Epoch},
        ylabel={Q-Factor [dB]},
        ylabel shift = {-5},
        grid=both,  
        xmin=0, xmax=500,
    	ymin=0, ymax=5.5,
        legend style={legend pos=south west, legend cell align=left,fill=white, fill opacity=0.6, draw opacity=1,text opacity=1},
    	grid style={dashed}]
        ]
    \addplot[color=red,very thick]    coordinates {(1, 4.37)(2, 4.68)(3, 4.81)(4, 4.87)(5, 4.91)(6, 4.93)(7, 4.94)(8, 4.96)(9, 4.95)(10, 4.99)(11, 5.0)(12, 5.02)(13, 5.02)(14, 5.03)(15, 5.01)(16, 5.01)(17, 5.05)(18, 5.01)(19, 5.02)(20, 5.04)(21, 5.07)(22, 5.05)(23, 5.03)(24, 5.06)(25, 5.07)(26, 5.06)(27, 5.08)(28, 5.06)(29, 5.04)(30, 5.07)(31, 5.06)(32, 5.08)(33, 5.06)(34, 5.08)(35, 5.06)(36, 5.06)(37, 5.08)(38, 5.08)(39, 5.08)(40, 5.04)(41, 5.09)(42, 5.08)(43, 5.07)(44, 5.07)(45, 5.09)(46, 5.07)(47, 5.08)(48, 5.07)(49, 5.08)(50, 5.08)(51, 5.07)(52, 5.03)(53, 5.07)(54, 5.06)(55, 5.07)(56, 5.09)(57, 5.06)(58, 5.06)(59, 5.08)(60, 5.09)(61, 5.09)(62, 5.09)(63, 5.1)(64, 5.06)(65, 5.09)(66, 5.07)(67, 5.09)(68, 5.1)(69, 5.08)(70, 5.07)(71, 5.09)(72, 5.08)(73, 5.07)(74, 5.07)(75, 5.06)(76, 5.08)(77, 5.07)(78, 5.07)(79, 5.08)(80, 5.08)(81, 5.08)(82, 5.07)(83, 5.06)(84, 5.09)(85, 5.08)(86, 5.07)(87, 5.08)(88, 5.08)(89, 5.01)(90, 5.09)(91, 5.07)(92, 5.08)(93, 5.09)(94, 5.07)(95, 5.05)(96, 5.02)(97, 5.08)(98, 5.07)(99, 5.09)(100, 5.1)(101, 5.08)(102, 5.09)(103, 5.08)(104, 5.06)(105, 5.09)(106, 5.09)(107, 5.09)(108, 5.07)(109, 5.09)(110, 5.08)(111, 5.08)(112, 5.06)(113, 5.08)(114, 5.08)(115, 5.09)(116, 5.08)(117, 5.09)(118, 5.05)(119, 5.09)(120, 5.05)(121, 5.08)(122, 5.04)(123, 5.1)(124, 5.09)(125, 5.09)(126, 5.09)(127, 5.06)(128, 5.09)(129, 5.09)(130, 5.09)(131, 5.07)(132, 5.09)(133, 5.09)(134, 5.05)(135, 5.08)(136, 5.06)(137, 5.1)(138, 5.08)(139, 5.09)(140, 5.1)(141, 5.06)(142, 5.08)(143, 5.06)(144, 5.08)(145, 5.08)(146, 5.1)(147, 5.08)(148, 5.08)(149, 5.1)(150, 5.09)(151, 5.07)(152, 5.1)(153, 5.09)(154, 5.09)(155, 5.08)(156, 5.07)(157, 5.08)(158, 5.09)(159, 5.1)(160, 5.1)(161, 5.08)(162, 5.09)(163, 5.09)(164, 5.05)(165, 5.06)(166, 5.09)(167, 5.09)(168, 5.08)(169, 5.09)(170, 5.08)(171, 5.1)(172, 5.06)(173, 5.08)(174, 5.08)(175, 5.09)(176, 5.08)(177, 5.1)(178, 5.1)(179, 5.08)(180, 5.07)(181, 5.1)(182, 5.09)(183, 5.09)(184, 5.08)(185, 5.06)(186, 5.09)(187, 5.06)(188, 5.1)(189, 5.08)(190, 5.08)(191, 5.09)(192, 5.06)(193, 5.08)(194, 5.08)(195, 5.09)(196, 5.06)(197, 5.08)(198, 5.08)(199, 5.09)(200, 5.08)(201, 5.07)(202, 5.08)(203, 5.07)(204, 5.07)(205, 5.07)(206, 5.08)(207, 5.09)(208, 5.07)(209, 5.08)(210, 5.08)(211, 5.09)(212, 5.09)(213, 5.07)(214, 5.08)(215, 5.07)(216, 5.08)(217, 5.07)(218, 5.09)(219, 5.1)(220, 5.06)(221, 5.09)(222, 5.05)(223, 5.07)(224, 5.1)(225, 5.09)(226, 5.09)(227, 5.06)(228, 5.09)(229, 5.09)(230, 5.07)(231, 5.08)(232, 5.08)(233, 5.09)(234, 5.05)(235, 5.08)(236, 5.09)(237, 5.1)(238, 5.08)(239, 5.08)(240, 5.09)(241, 5.07)(242, 5.06)(243, 5.09)(244, 5.08)(245, 5.07)(246, 5.08)(247, 5.09)(248, 5.07)(249, 5.09)(250, 5.05)(251, 5.09)(252, 5.07)(253, 5.07)(254, 5.09)(255, 5.08)(256, 5.1)(257, 5.09)(258, 5.1)(259, 5.08)(260, 5.1)(261, 5.02)(262, 5.1)(263, 5.07)(264, 5.07)(265, 5.08)(266, 5.07)(267, 5.1)(268, 4.99)(269, 5.09)(270, 5.09)(271, 5.1)(272, 5.08)(273, 5.1)(274, 5.08)(275, 5.08)(276, 5.08)(277, 5.05)(278, 5.1)(279, 5.07)(280, 5.09)(281, 5.08)(282, 5.08)(283, 5.06)(284, 5.08)(285, 5.1)(286, 5.08)(287, 5.1)(288, 5.1)(289, 5.09)(290, 5.07)(291, 5.07)(292, 5.07)(293, 5.09)(294, 5.08)(295, 5.07)(296, 5.07)(297, 5.08)(298, 5.09)(299, 5.09)(300, 5.08)(301, 5.08)(302, 5.09)(303, 5.06)(304, 5.1)(305, 5.07)(306, 5.08)(307, 5.05)(308, 5.1)(309, 5.07)(310, 5.09)(311, 5.1)(312, 5.1)(313, 5.1)(314, 5.09)(315, 5.08)(316, 5.09)(317, 5.07)(318, 5.08)(319, 5.05)(320, 5.05)(321, 5.06)(322, 5.06)(323, 5.09)(324, 5.09)(325, 5.09)(326, 5.08)(327, 5.1)(328, 5.1)(329, 5.09)(330, 5.1)(331, 5.08)(332, 5.09)(333, 5.09)(334, 5.08)(335, 5.07)(336, 5.09)(337, 5.08)(338, 5.09)(339, 5.1)(340, 5.09)(341, 5.06)(342, 5.09)(343, 5.09)(344, 5.05)(345, 5.08)(346, 5.1)(347, 5.08)(348, 5.08)(349, 5.09)(350, 5.08)(351, 5.08)(352, 5.07)(353, 5.06)(354, 5.04)(355, 5.08)(356, 5.09)(357, 5.09)(358, 5.11)(359, 5.08)(360, 5.07)(361, 5.11)(362, 5.08)(363, 5.09)(364, 5.04)(365, 5.08)(366, 5.1)(367, 5.07)(368, 5.07)(369, 5.05)(370, 5.06)(371, 5.1)(372, 5.08)(373, 5.1)(374, 5.05)(375, 5.09)(376, 5.06)(377, 5.09)(378, 5.09)(379, 4.98)(380, 5.09)(381, 5.05)(382, 5.1)(383, 5.05)(384, 5.08)(385, 5.09)(386, 5.06)(387, 5.06)(388, 5.07)(389, 5.08)(390, 5.07)(391, 5.08)(392, 5.07)(393, 5.07)(394, 5.07)(395, 5.09)(396, 5.06)(397, 5.09)(398, 5.09)(399, 5.05)(400, 5.08)(401, 5.08)(402, 5.08)(403, 5.09)(404, 5.08)(405, 5.09)(406, 5.08)(407, 5.06)(408, 5.06)(409, 5.04)(410, 5.06)(411, 5.05)(412, 5.09)(413, 5.09)(414, 5.08)(415, 5.07)(416, 5.08)(417, 5.09)(418, 5.07)(419, 5.04)(420, 5.07)(421, 5.06)(422, 5.09)(423, 5.09)(424, 5.09)(425, 5.06)(426, 5.09)(427, 5.05)(428, 5.09)(429, 5.09)(430, 5.08)(431, 5.1)(432, 5.08)(433, 5.09)(434, 5.1)(435, 5.09)(436, 5.09)(437, 5.1)(438, 5.09)(439, 5.07)(440, 5.08)(441, 5.08)(442, 5.07)(443, 5.08)(444, 5.06)(445, 5.1)(446, 5.05)(447, 5.1)(448, 5.08)(449, 5.08)(450, 5.07)(451, 5.09)(452, 5.09)(453, 5.09)(454, 5.09)(455, 5.07)(456, 5.04)(457, 5.03)(458, 5.07)(459, 5.09)(460, 5.09)(461, 5.07)(462, 5.08)(463, 5.1)(464, 5.08)(465, 5.07)(466, 5.05)(467, 5.09)(468, 5.09)(469, 5.08)(470, 5.08)(471, 5.06)(472, 5.09)(473, 5.1)(474, 5.08)(475, 5.09)(476, 5.06)(477, 5.06)(478, 5.07)(479, 5.07)(480, 5.1)(481, 5.08)(482, 5.09)(483, 5.09)(484, 5.03)(485, 5.08)(486, 5.05)(487, 5.06)(488, 5.04)(489, 5.09)(490, 5.09)(491, 5.06)(492, 5.1)(493, 5.08)(494, 5.07)(495, 5.09)(496, 5.09)(497, 5.09)(498, 5.07)(499, 5.08)(500, 5.08)
    };
    \addlegendentry{Taylor ($3^\text{rd}$ order)};
    
    \addplot[color=blue,very thick]    coordinates {
    (1, 3.86)(2, 4.25)(3, 4.42)(4, 4.53)(5, 4.6)(6, 4.66)(7, 4.68)(8, 4.72)(9, 4.72)(10, 4.75)(11, 4.8)(12, 4.8)(13, 4.82)(14, 4.83)(15, 4.84)(16, 4.85)(17, 4.85)(18, 4.86)(19, 4.86)(20, 4.88)(21, 4.89)(22, 4.88)(23, 4.89)(24, 4.89)(25, 4.91)(26, 4.91)(27, 4.91)(28, 4.92)(29, 4.91)(30, 4.84)(31, 4.94)(32, 4.94)(33, 4.94)(34, 4.94)(35, 4.94)(36, 4.94)(37, 4.96)(38, 4.94)(39, 4.93)(40, 4.92)(41, 4.96)(42, 4.96)(43, 4.95)(44, 4.96)(45, 4.97)(46, 4.93)(47, 4.96)(48, 4.95)(49, 4.96)(50, 4.97)(51, 4.95)(52, 4.95)(53, 4.97)(54, 4.93)(55, 4.95)(56, 4.98)(57, 4.95)(58, 4.94)(59, 4.97)(60, 4.97)(61, 5.0)(62, 4.98)(63, 4.99)(64, 4.98)(65, 4.99)(66, 4.98)(67, 4.98)(68, 4.98)(69, 4.98)(70, 4.97)(71, 4.97)(72, 4.99)(73, 4.99)(74, 4.97)(75, 4.98)(76, 5.0)(77, 5.0)(78, 4.98)(79, 4.95)(80, 4.99)(81, 4.94)(82, 4.99)(83, 4.96)(84, 4.98)(85, 4.98)(86, 4.98)(87, 5.01)(88, 4.99)(89, 4.97)(90, 4.99)(91, 4.99)(92, 5.0)(93, 4.99)(94, 5.0)(95, 4.95)(96, 4.94)(97, 5.0)(98, 4.98)(99, 4.98)(100, 5.0)(101, 5.0)(102, 4.99)(103, 5.0)(104, 5.01)(105, 5.0)(106, 5.0)(107, 5.01)(108, 4.98)(109, 5.01)(110, 4.99)(111, 5.0)(112, 4.99)(113, 4.98)(114, 4.99)(115, 5.0)(116, 4.99)(117, 5.01)(118, 4.96)(119, 5.0)(120, 4.98)(121, 5.02)(122, 4.96)(123, 5.02)(124, 4.97)(125, 5.0)(126, 5.02)(127, 5.0)(128, 5.01)(129, 5.0)(130, 5.02)(131, 5.0)(132, 5.01)(133, 5.01)(134, 4.93)(135, 5.01)(136, 4.98)(137, 5.01)(138, 4.95)(139, 5.0)(140, 5.01)(141, 5.01)(142, 4.99)(143, 4.98)(144, 4.96)(145, 5.01)(146, 5.02)(147, 5.02)(148, 5.02)(149, 5.01)(150, 4.99)(151, 5.01)(152, 5.02)(153, 5.01)(154, 5.02)(155, 5.01)(156, 5.02)(157, 5.0)(158, 5.0)(159, 5.02)(160, 5.01)(161, 5.02)(162, 5.02)(163, 5.01)(164, 5.01)(165, 4.99)(166, 5.01)(167, 5.03)(168, 5.01)(169, 5.03)(170, 5.01)(171, 5.01)(172, 5.0)(173, 5.02)(174, 5.02)(175, 5.01)(176, 5.0)(177, 5.02)(178, 5.01)(179, 5.02)(180, 4.97)(181, 5.02)(182, 5.02)(183, 5.03)(184, 5.0)(185, 5.02)(186, 5.03)(187, 5.0)(188, 5.03)(189, 5.02)(190, 5.02)(191, 5.0)(192, 5.01)(193, 5.02)(194, 5.02)(195, 5.04)(196, 5.01)(197, 5.03)(198, 5.03)(199, 5.03)(200, 5.02)(201, 4.99)(202, 5.02)(203, 5.02)(204, 5.02)(205, 4.99)(206, 5.02)(207, 5.01)(208, 5.02)(209, 5.02)(210, 4.98)(211, 5.02)(212, 5.03)(213, 4.99)(214, 4.98)(215, 5.01)(216, 5.01)(217, 5.03)(218, 5.02)(219, 5.03)(220, 4.99)(221, 5.03)(222, 5.03)(223, 4.98)(224, 5.02)(225, 5.02)(226, 5.03)(227, 5.0)(228, 5.03)(229, 5.03)(230, 5.0)(231, 5.03)(232, 5.01)(233, 5.03)(234, 4.98)(235, 5.03)(236, 5.03)(237, 5.03)(238, 5.02)(239, 5.02)(240, 5.02)(241, 4.99)(242, 5.04)(243, 5.01)(244, 5.03)(245, 5.03)(246, 5.0)(247, 5.03)(248, 5.03)(249, 5.02)(250, 4.99)(251, 5.02)(252, 5.02)(253, 4.99)(254, 5.02)(255, 5.01)(256, 5.04)(257, 5.02)(258, 5.03)(259, 5.02)(260, 5.03)(261, 5.02)(262, 5.0)(263, 5.02)(264, 5.03)(265, 5.01)(266, 5.02)(267, 5.03)(268, 4.96)(269, 5.03)(270, 5.02)(271, 5.02)(272, 5.01)(273, 5.03)(274, 5.03)(275, 5.03)(276, 5.03)(277, 5.02)(278, 5.03)(279, 5.03)(280, 5.02)(281, 5.02)(282, 5.02)(283, 5.03)(284, 5.03)(285, 5.03)(286, 5.02)(287, 5.03)(288, 5.04)(289, 5.02)(290, 5.01)(291, 5.02)(292, 5.02)(293, 5.03)(294, 5.02)(295, 5.0)(296, 5.0)(297, 5.02)(298, 5.04)(299, 5.03)(300, 5.01)(301, 5.02)(302, 5.02)(303, 5.0)(304, 5.03)(305, 5.02)(306, 5.03)(307, 5.02)(308, 5.04)(309, 5.01)(310, 5.02)(311, 5.03)(312, 5.01)(313, 5.03)(314, 5.04)(315, 5.02)(316, 5.03)(317, 5.03)(318, 5.03)(319, 5.01)(320, 5.0)(321, 5.02)(322, 5.01)(323, 5.02)(324, 5.02)(325, 5.03)(326, 5.02)(327, 5.04)(328, 5.02)(329, 5.04)(330, 5.04)(331, 5.01)(332, 5.03)(333, 5.03)(334, 5.0)(335, 5.03)(336, 5.02)(337, 5.02)(338, 5.02)(339, 5.03)(340, 5.04)(341, 5.02)(342, 5.02)(343, 5.03)(344, 4.99)(345, 5.02)(346, 5.02)(347, 5.03)(348, 5.02)(349, 5.02)(350, 5.04)(351, 5.02)(352, 5.03)(353, 5.0)(354, 4.99)(355, 5.04)(356, 5.04)(357, 5.03)(358, 5.02)(359, 5.02)(360, 5.01)(361, 5.02)(362, 5.02)(363, 5.03)(364, 4.97)(365, 5.03)(366, 5.03)(367, 5.03)(368, 5.03)(369, 4.99)(370, 5.0)(371, 5.04)(372, 5.02)(373, 5.04)(374, 5.01)(375, 5.02)(376, 5.0)(377, 5.02)(378, 5.03)(379, 5.0)(380, 5.03)(381, 4.97)(382, 5.04)(383, 5.0)(384, 5.02)(385, 5.02)(386, 4.99)(387, 5.0)(388, 5.0)(389, 5.01)(390, 5.03)(391, 5.03)(392, 5.02)(393, 5.01)(394, 5.01)(395, 5.03)(396, 5.0)(397, 5.03)(398, 5.04)(399, 5.01)(400, 5.03)(401, 5.02)(402, 5.03)(403, 5.03)(404, 5.02)(405, 5.03)(406, 5.03)(407, 5.01)(408, 5.02)(409, 4.99)(410, 5.02)(411, 5.01)(412, 5.04)(413, 5.04)(414, 5.04)(415, 5.02)(416, 5.03)(417, 5.02)(418, 5.04)(419, 4.98)(420, 5.02)(421, 5.01)(422, 5.01)(423, 5.03)(424, 5.03)(425, 5.01)(426, 5.04)(427, 5.0)(428, 5.05)(429, 5.03)(430, 5.02)(431, 5.02)(432, 5.01)(433, 5.04)(434, 5.04)(435, 5.03)(436, 5.04)(437, 5.04)(438, 5.04)(439, 5.03)(440, 5.03)(441, 5.02)(442, 5.01)(443, 5.03)(444, 5.02)(445, 5.05)(446, 5.01)(447, 5.04)(448, 5.03)(449, 5.02)(450, 5.03)(451, 5.04)(452, 5.03)(453, 5.04)(454, 5.04)(455, 5.0)(456, 4.99)(457, 5.02)(458, 5.03)(459, 5.04)(460, 5.03)(461, 5.02)(462, 5.02)(463, 5.04)(464, 5.04)(465, 5.02)(466, 5.01)(467, 5.03)(468, 5.04)(469, 5.02)(470, 5.03)(471, 5.02)(472, 5.01)(473, 5.03)(474, 5.03)(475, 5.03)(476, 5.04)(477, 5.02)(478, 5.01)(479, 5.0)(480, 5.04)(481, 5.03)(482, 5.04)(483, 5.03)(484, 5.04)(485, 5.04)(486, 5.01)(487, 5.01)(488, 5.01)(489, 5.04)(490, 5.05)(491, 5.03)(492, 5.04)(493, 5.04)(494, 5.03)(495, 5.03)(496, 5.03)(497, 5.03)(498, 5.03)(499, 5.03)(500, 5.04)
    };\addlegendentry{PWL ($3$ segments)};
    
    \addplot[color=violet,very thick]    coordinates {(1, 2.86)(2, 3.04)(3, 3.2)(4, 3.34)(5, 3.45)(6, 3.53)(7, 3.6)(8, 3.67)(9, 3.74)(10, 3.82)(11, 3.86)(12, 3.9)(13, 3.95)(14, 3.94)(15, 3.98)(16, 4.01)(17, 4.02)(18, 4.05)(19, 4.1)(20, 4.1)(21, 4.13)(22, 4.18)(23, 4.2)(24, 4.21)(25, 4.21)(26, 4.24)(27, 4.26)(28, 4.26)(29, 4.3)(30, 4.3)(31, 4.32)(32, 4.33)(33, 4.34)(34, 4.37)(35, 4.37)(36, 4.36)(37, 4.41)(38, 4.39)(39, 4.41)(40, 4.42)(41, 4.43)(42, 4.4)(43, 4.44)(44, 4.44)(45, 4.45)(46, 4.44)(47, 4.45)(48, 4.45)(49, 4.39)(50, 4.43)(51, 4.45)(52, 4.45)(53, 4.44)(54, 4.48)(55, 4.43)(56, 4.46)(57, 3.54)(58, 3.88)(59, 3.99)(60, 4.05)(61, 4.11)(62, 4.17)(63, 4.19)(64, 4.21)(65, 4.26)(66, 4.29)(67, 4.29)(68, 4.31)(69, 4.32)(70, 4.33)(71, 4.36)(72, 4.38)(73, 4.37)(74, 4.38)(75, 4.39)(76, 4.38)(77, 4.38)(78, 4.41)(79, 4.4)(80, 4.41)(81, 4.43)(82, 4.43)(83, 4.42)(84, 4.42)(85, 4.43)(86, 4.44)(87, 4.41)(88, 4.43)(89, 4.42)(90, 4.45)(91, 4.43)(92, 4.44)(93, 4.43)(94, 4.43)(95, 4.42)(96, 4.44)(97, 4.43)(98, 4.43)(99, 4.44)(100, 4.42)(101, 2.34)(102, 4.21)(103, 4.33)(104, 4.38)(105, 4.41)(106, 4.41)(107, 4.42)(108, 4.4)(109, 4.42)(110, 0.8)(111, 3.42)(112, 3.56)(113, 3.75)(114, 3.87)(115, 3.95)(116, 4.03)(117, 4.02)(118, 4.14)(119, 4.21)(120, 4.27)(121, 4.3)(122, 4.33)(123, 4.34)(124, 4.36)(125, 4.37)(126, 4.4)(127, 4.39)(128, 4.37)(129, 4.29)(130, 4.18)(131, 4.13)(132, 4.22)(133, 3.26)(134, 1.82)(135, 2.27)(136, 2.49)(137, 2.62)(138, 2.68)(139, 2.82)(140, 2.95)(141, 3.04)(142, 3.11)(143, 3.18)(144, 3.24)(145, 3.28)(146, 3.33)(147, 3.37)(148, 3.41)(149, 3.43)(150, 3.48)(151, 3.5)(152, 3.53)(153, 3.57)(154, 3.58)(155, 3.61)(156, 3.64)(157, 3.66)(158, 3.67)(159, 3.69)(160, 3.7)(161, 3.74)(162, 3.75)(163, 3.76)(164, 3.8)(165, 3.82)(166, 3.82)(167, 3.85)(168, 3.87)(169, 3.88)(170, 3.91)(171, 3.92)(172, 3.93)(173, 3.95)(174, 3.95)(175, 3.96)(176, 3.98)(177, 4.01)(178, 4.01)(179, 4.03)(180, 4.05)(181, 4.06)(182, 4.06)(183, 4.09)(184, 4.09)(185, 4.11)(186, 4.13)(187, 4.13)(188, 4.14)(189, 4.13)(190, 4.13)(191, 4.12)(192, 4.12)(193, 4.1)(194, 4.13)(195, 4.13)(196, 4.1)(197, 4.16)(198, 4.14)(199, 4.17)(200, 4.09)(201, 4.13)(202, 4.16)(203, 4.02)(204, 4.04)(205, 4.14)(206, 4.18)(207, 4.13)(208, 4.18)(209, 4.13)(210, 4.13)(211, 4.17)(212, 4.19)(213, 4.17)(214, 4.14)(215, 3.8)(216, 4.12)(217, 4.2)(218, 4.15)(219, 4.16)(220, 4.02)(221, 4.1)(222, 4.1)(223, 4.19)(224, 4.18)(225, 4.07)(226, 3.92)(227, 3.83)(228, 3.8)(229, 3.86)(230, 3.83)(231, 3.77)(232, 3.78)(233, 3.7)(234, 3.54)(235, 3.58)(236, 3.58)(237, 3.57)(238, 3.36)(239, 3.47)(240, 3.54)(241, 3.56)(242, 3.58)(243, 3.6)(244, 3.61)(245, 3.63)(246, 3.64)(247, 3.63)(248, 3.66)(249, 3.67)(250, 3.67)(251, 3.66)(252, 3.68)(253, 3.68)(254, 3.67)(255, 3.67)(256, 3.68)(257, 3.69)(258, 3.68)(259, 3.68)(260, 3.69)(261, 3.69)(262, 3.68)(263, 3.68)(264, 3.69)(265, 3.69)(266, 3.7)(267, 3.7)(268, 3.69)(269, 3.66)(270, 3.7)(271, 3.69)(272, 3.69)(273, 3.68)(274, 3.68)(275, 3.69)(276, 3.69)(277, 3.66)(278, 3.66)(279, 2.1)(280, 2.42)(281, 2.54)(282, 2.63)(283, 2.67)(284, 2.72)(285, 2.77)(286, 2.8)(287, 2.82)(288, 2.84)(289, 2.86)(290, 2.88)(291, 2.9)(292, 2.91)(293, 2.93)(294, 2.95)(295, 2.96)(296, 2.98)(297, 2.97)(298, 2.99)(299, 3.0)(300, 3.02)(301, 3.02)(302, 3.04)(303, 3.05)(304, 3.06)(305, 3.07)(306, 3.09)(307, 3.09)(308, 3.11)(309, 3.12)(310, 3.12)(311, 3.12)(312, 3.14)(313, 3.17)(314, 3.18)(315, 3.18)(316, 3.21)(317, 3.22)(318, 3.24)(319, 3.26)(320, 3.28)(321, 3.24)(322, 3.31)(323, 3.33)(324, 3.34)(325, 2.73)(326, 2.96)(327, 3.01)(328, 3.02)(329, 3.03)(330, 2.99)(331, 3.02)(332, 3.03)(333, 3.04)(334, 3.05)(335, 3.04)(336, 3.04)(337, 3.05)(338, 3.06)(339, 3.07)(340, 3.05)(341, 3.06)(342, 3.06)(343, 3.06)(344, 3.07)(345, 3.08)(346, 3.07)(347, 3.08)(348, 3.08)(349, 3.08)(350, 3.07)(351, 3.08)(352, 3.09)(353, 3.09)(354, 3.07)(355, 3.1)(356, 3.09)(357, 3.09)(358, 3.1)(359, 3.09)(360, 3.09)(361, 3.1)(362, 3.09)(363, 3.09)(364, 3.08)(365, 3.09)(366, 3.05)(367, 3.13)(368, 3.15)(369, 3.16)(370, 3.16)(371, 3.17)(372, 3.18)(373, 3.2)(374, 3.19)(375, 3.19)(376, 3.2)(377, 3.2)(378, 3.2)(379, 3.19)(380, 3.19)(381, 3.21)(382, 3.22)(383, 3.2)(384, 3.2)(385, 3.21)(386, 3.19)(387, 3.19)(388, 3.21)(389, 3.2)(390, 3.2)(391, 3.21)(392, 3.21)(393, 3.2)(394, 3.21)(395, 3.2)(396, 3.19)(397, 3.13)(398, 2.87)(399, 3.03)(400, 3.08)(401, 3.09)(402, 3.1)(403, 3.13)(404, 3.14)(405, 3.14)(406, 3.15)(407, 3.16)(408, 3.17)(409, 3.18)(410, 3.18)(411, 3.2)(412, 3.2)(413, 3.2)(414, 3.21)(415, 3.22)(416, 3.22)(417, 3.23)(418, 3.24)(419, 3.24)(420, 3.26)(421, 3.26)(422, 3.27)(423, 3.28)(424, 3.28)(425, 3.29)(426, 3.3)(427, 3.31)(428, 2.9)(429, 3.03)(430, 3.08)(431, 3.09)(432, 3.12)(433, 3.13)(434, 3.13)(435, 3.16)(436, 3.16)(437, 3.18)(438, 3.17)(439, 3.2)(440, 3.2)(441, 3.22)(442, 3.23)(443, 3.23)(444, 3.25)(445, 3.24)(446, 3.26)(447, 3.26)(448, 3.26)(449, 3.27)(450, 3.26)(451, 3.25)(452, 3.24)(453, 2.96)(454, 2.99)(455, 2.76)(456, 2.86)(457, 2.9)(458, 2.93)(459, 2.95)(460, 2.96)(461, 2.97)(462, 2.98)(463, 2.99)(464, 3.0)(465, 3.01)(466, 3.0)(467, 3.02)(468, 3.02)(469, 3.03)(470, 3.03)(471, 3.02)(472, 3.03)(473, 3.01)(474, 3.05)(475, 3.05)(476, 3.06)(477, 3.06)(478, 3.07)(479, 3.08)(480, 3.07)(481, 3.06)(482, 3.08)(483, 3.08)(484, 3.09)(485, 3.1)(486, 3.1)(487, 3.09)(488, 3.11)(489, 3.11)(490, 3.08)(491, 3.1)(492, 3.11)(493, 3.1)(494, 3.08)(495, 3.11)(496, 3.1)(497, 3.11)(498, 3.11)(499, 3.12)(500, 3.1)};
    \addlegendentry{LUT ($n_\text{bit}=7$)};
    \end{axis}
    \end{tikzpicture}